\documentclass{aa}  
\usepackage[switch]{lineno}%\pagewiselinenumbers
\usepackage[dvipsnames]{xcolor}
\usepackage{colortbl}
\usepackage{graphicx}
\usepackage{txfonts}
 \usepackage{hyperref}
 \hypersetup{
     colorlinks=true,
     linkcolor=[RGB]{73,148,204},
     filecolor=cyan,
     citecolor=[RGB]{73,148,204},      
     urlcolor=[RGB]{73,148,204},
     }
\begin{document} 

   \title{Relationship between the $\gamma$-ray variability and the parsec-scale jet in the blazar 3C 454.3}
   \author{E. Palafox\inst{1,2},%\fnmsep\thanks{eva.palafox@gmail.com},
          V. M. Patiño-Álvarez\inst{1,2},
          %\and
          V. Chavushyan\inst{1},
          A. Lobanov\inst{2},
          S. A. Dzib\inst{2},
          \and
          A. Zensus\inst{2} 
          }
          
\institute{Instituto Nacional de Astrofísica, Óptica y Electrónica, Luis Enrique Erro $\#$1, Tonantzintla, Puebla 72840, México\\
\email{epalafox.astro@gmail.com}
\and Max-Planck-Institute für Radioastronomie, Auf dem Hügel 69, 53121 Bonn, Germany
          }
             
\date{Version: \today; Received September 16, 2024; accepted February 17, 2025 }
   
% \abstract{}{}{}{}{} 
% 5 {} token are mandatory 
  \abstract
   {
   The flat spectrum radio quasar 3C 454.3 is known for its high variability across the electromagnetic spectrum, showing structural and flux variability in its parsec-scale jet and correlated variability among frequency bands. 
   }
   {
   This study aims to identify the structure, dynamics, and radiative processes common to the innermost regions of the blazar 3C 454.3. We investigate whether any jet component can be associated with $\gamma$-ray emission and variability. 
   Additionally, we compare the flux variability of the parsec-scale jet components found in the VLBA observations to the variability in the $\gamma$-ray emission. 
   }
   {   
    We analyzed the relationship between the variable $\gamma$-ray emission and parsec-scale jet properties in 3C 454.3 by combining $\gamma$-ray data spanning twelve years with contemporaneous VLBA multi-epoch images at 15 and 43 GHz.
    We conducted Spearman’s rank correlation tests to determine if the flux variability of any jet component is associated with $\gamma$-ray variability. 
   }
   {
    The core emission at 43 and 15 GHz strongly correlates with $\gamma$-ray emission. 
    The 43 GHz core (Q0) contributes around 37$\%$ of the observed $\gamma$-ray variability, while the 15 GHz core (K0) accounts for 30$\%$. A quasi-stationary component at 43 GHz, at a projected distance of 4.6 pc, correlates with the $\gamma$-ray flux, accounting for 20$\%$ of its emission between 2016 and 2021. We found a mobile component (Q3 between 2010.18 and 2011.16) at 43 GHz with a projected distance between 0.8 and 2.3 pc and an apparent velocity of $\beta_{app} = 9.9 \pm 1.1$ c that accounts for approximately 28$\%$ of the $\gamma$-ray emission. The observed simultaneous variability in emission regions beyond the central parsec strongly suggests synchrotron self-Compton as the primary mechanism for $\gamma$-ray production in these regions.
   }
   {
    Our findings demonstrate the existence of multiple $\gamma$-ray emission regions within the blazar jet but also suggest that some of these regions are non-stationary over time. Additionally, our study pinpoints the exact locations of these emission regions within the blazar itself. These results are valuable for theoretical models and for gaining a deeper understanding of the complex nature of blazars.
   }

   \keywords{ $\gamma$ rays: galaxies --- radiation mechanisms: non-thermal --- quasars: individual: 3C 454.3 --- galaxies: active --- techniques: high angular resolution}
%-------------------------------------------------------------------
   \titlerunning{$\gamma$-ray variability and the parsec-scale jet in the blazar 3C 454.3}
    \authorrunning{Palafox, E., et al.} 
   \maketitle
%-------------------------------------------------------------------
\section{Introduction}
%-------------------------------------------------------------------
%-------------------------------------------------------------------
\begin{figure*}[ht!]
    \centering
    \includegraphics[width=0.9\textwidth]{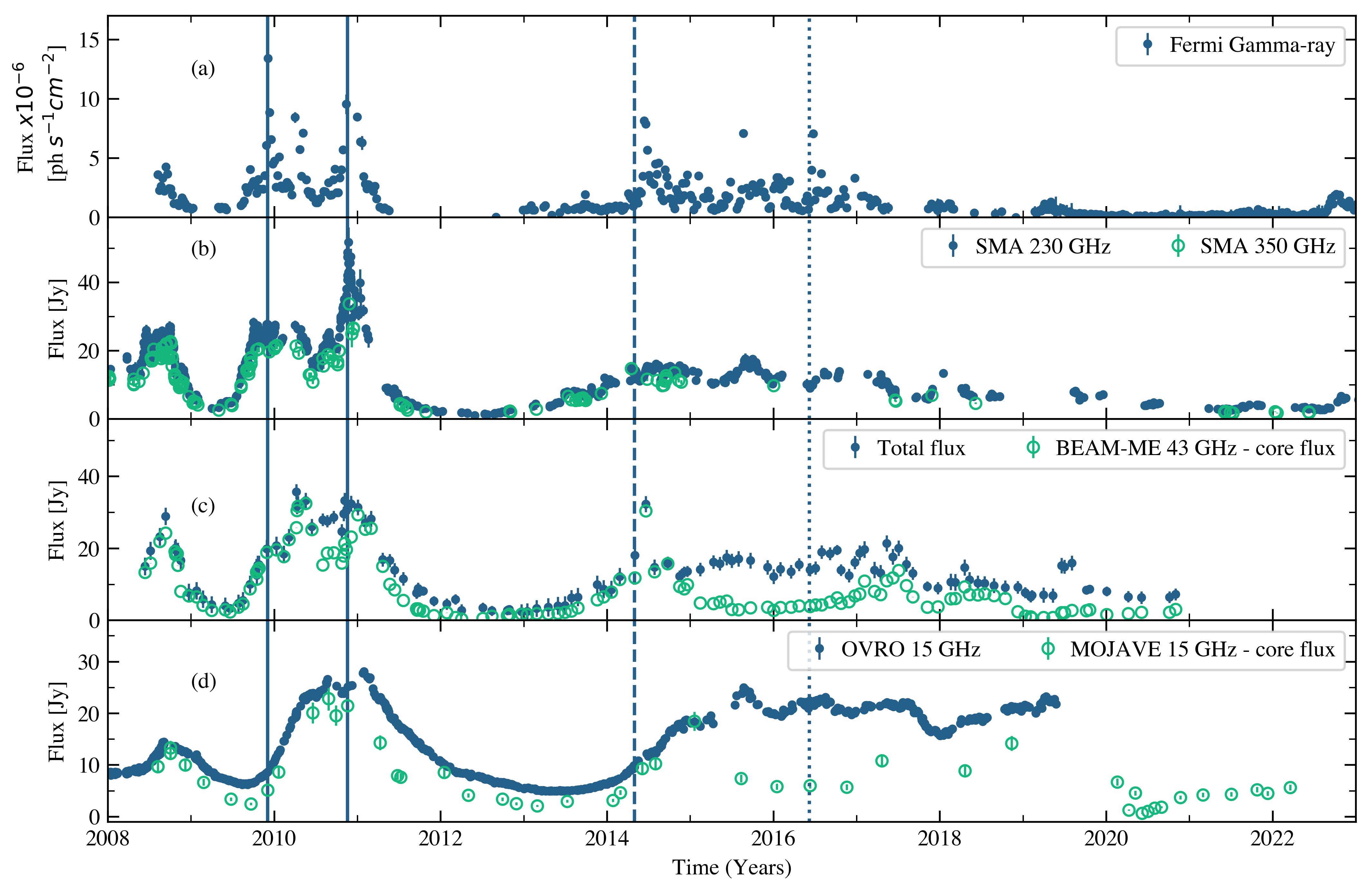}
    \caption{Multi-wavelength light curves of 3C 454.3. Each panel shows the blazar emission at a different wavelength (indicated by the label and color). The data source is specified within each panel. The solid vertical lines mark $\gamma$-ray flare events that appear to coincide with the passage of jet components K09 and K10 through the VLBI core \citep{2013ApJ...773..147J}. The dashed vertical line indicates the ejection of a relativistic component from the radio core (K14; \citealt{2020ApJ...902...61L}). 
    The dotted vertical line represents the time at which a knot crossed the 43 GHz core (K16; \citealt{2019ApJ...875...15W}).}
    \label{fig:lcurve}
\end{figure*}
%-------------------------------------------------------------------

%\textcolor{red}{General notes: I have edited to US English spelling and grammar conventions. Please review the following style notes, as they explain some of the corrections and provide guidance for further correction: (A.) A\&A uses the past tense to describe specific methods used in a paper and the present tense to describe general methods as well as findings, including the findings of recent papers (within the past ten or so years). See Sect. 6 of the language guide https://www.aanda.org/for-authors/language-editing/6-verb-tenses. (B.) Acronyms/Abbreviations - Acronyms/abbreviations should be introduced upon first use, and thereafter the acronym/abbreviation should be used unless it’s at the beginning of a sentence of a new paragraph. Also note that if an abbreviation is not used, it does not need to be introduced. (C.) Numbers/Numerals - Numbers less than 11 that are not a measurement should be written out (i.e., three stars vs. 3 cm). (D.) Units - Units should be written out when they aren't directly accompanied by numerals.}
Blazars \citep{1980ARA&A..18..321A} are among the least common class of active galactic nuclei (AGN), and their emission is dominated by radiation from non–thermal plasma moving at a relativistic speed in a direction close to the line of sight. Given the small viewing angle of the jet (e.g., \citealp{2019NewAR..8701541H}), blazar emission shows relativistic effects such as beaming and Doppler boosting that cause the jet components to move with apparent superluminal speeds of up to ten times the speed of light (e.g., \citealp{2017ApJ...846...98J}). 
Although blazars make up the majority of discrete $\gamma$-ray sources seen by the Fermi Large Area Telescope (\citealp{2022ApJS..260...53A}), such questions as what the mechanisms of $\gamma$-ray production are, where they occur, and if there can be different mechanisms operating even in the same object remain open. 
The spectral energy distribution of blazars shows two peaks \citep{1989MNRAS.236..341G, 1993ApJ...416..458D, 1994ApJ...421..153S}. The first one is due to synchrotron emission, and the second one, at higher energies, is generally attributed to inverse Compton (IC) scattering.
The origin of the IC seed photons has been a topic of years-long debate. They could come from the accretion disk emission or from the broad-line region (BLR; \citealp{1994ApJ...421..153S}) or from the dusty torus \citep{2008ApJ...675...71S}, a process known as external IC (EIC) scattering. Alternatively, the seed photons can originate from synchrotron emission produced in the jet itself, a mechanism called synchrotron self-Compton (SSC, \citealp{1992ApJ...397L...5M, 1996ApJ...461..657B}). Hadronic mechanisms for producing the second peak have also been proposed (e.g., \citealp{1993A&A...269...67M}). 

% -------------
The correlated variability between different wavelengths can help locate the $\gamma$-ray emission region within a jet. However, Very Long Baseline Interferometry (VLBI) offers an even more powerful tool. VLBI observations at millimeter and radio wavelengths achieve the highest spatial resolution, thus allowing the exact location, or "blazar region," where $\gamma$-rays are produced to be identified (see review by \citealp{2017A&ARv..25....4B}).
Several studies have provided valuable insights into the location and physical processes responsible for $\gamma$-ray emission in AGN. For instance, \citep{2001ApJ...556..738J, 2003ApJ...590...95L, 2011A&A...532A.146L, 2014MNRAS.441.1899F} suggest that radio emission often occurs at the same time or after $\gamma$-rays. On the other hand, $\gamma$-ray emission has been detected from the blazar 3C 84 in multiple locations \citep{2018MNRAS.475..368H, 2023A&A...669A..32P} as well as from the blazars 1633+382 and 3C 345 \citep{2015MNRAS.452.1280R}. 
\citet{2013ApJ...763L..36L} linked variations in the Mg II $\lambda$2800 emission line with radio emission from superluminal jet material in 3C 454.3. This finding suggests the presence of broad emission line material well beyond the inner parsec close to the radio core. Similarly, \citet{2013ApJ...779..100I} provided evidence for a $\gamma-$emitting region interior to the broad-line region (BLR) during flares in 2009 and 2010 for the same object.
Recent VLBI observations have revealed even more complexity. \citet{2019A&A...630A..56P} studied 3C 279 and found, for the first time, evidence for a non-stationary $\gamma$-ray emission region located about 42 pc from the core. Their results suggest that the $\gamma$-ray emission may originate from different places within the jet.
Likewise, \citet{2020ApJ...891...68C} found evidence of BLR clouds at about 25 pc from the supermassive black hole in CTA 102.
Further supporting the dynamic nature of blazar jets, \citet{2024A&A...682A.154T} analyzed VLBI data of 3C 454.3 at high frequencies (43 GHz and 86 GHz) between 2013 and 2017. They observed four jet knots that seemingly disappeared upon reaching a quasi-stationary component at about 0.6 mas. The authors propose that this could be due to the jet folding back onto itself and thus redirecting the plasma toward our line of sight.

% ----------------- 
Notably, 3C 454.3 is known as one of the brightest sources in the $\gamma$-ray sky. 
Between 2009 and 2010, it presented a series of important $\gamma$-ray flares (\citealp{2010ApJ...721.1383A, 2011ApJ...736L..38V}). 
The late-2010 flare stands out as the brightest ever recorded at the time, reaching a peak flux of $F=8.5\pm0.5 \times10^{-5}$ photons cm$^{-2}$ s$^{-1}$ \citep{2011ApJ...733L..26A}.
This translates to an isotropic luminosity of approximately $10^{50}$ erg $s^{-1}$\citep{2021MNRAS.504.5074S}, solidifying its place among the brightest $\gamma$-ray sources known. 
Extensive research has confirmed a strong correlation between the $\gamma$-ray variability and radio frequencies, as well as other wavelengths (e.g., \citealp{2010ApJ...715..362J, 2013ApJ...763L..36L, 2020ApJ...902...61L, 2021ApJ...906....5A}). 
The passage of a plasma component down the jet is believed to be a primary driver for most $\gamma$-ray flares (e.g., \citealp{2010ApJ...710L.126M, 2013ApJ...773..147J, 2017ApJ...846...98J,  2019ApJ...875...15W}). 
In some cases, the flares coincide with the interaction of superluminal components within the jet, such as their passage through a quasi-stationary feature or the core (e.g., \citealp{2011ApJ...735L..10A}).
For instance, \citet{2013ApJ...773..147J} proposed that the 2009-2010 flares in 3C 454.3 might be linked to the emergence and propagation of two knots within the jet, designated as K09 and K10 and identified through analysis of monthly Very Long Baseline Array (VLBA) images.
Similarly, \citet{2019ApJ...875...15W} identified K16, a component ejected around four months before the $\gamma$-ray flare of 2016, which was also observed in optical. They suggest that K16 is likely responsible for the 2016 outburst when its lagging end moved past a standing shock in the core.
%This knot is likely responsible for a small-amplitude γ-ray outburst in 2016 March when the “head” of the knot moved past the standing shock, as well as the main 2016 outburst when the lagging end moved past the shock.
More recently \citet{2020ApJ...902...61L} reported that the $\gamma$-ray flare of 2014 might be the result of a traveling shock, the component K14, when it crossed a standing re-collimation shock observed as the radio core. 
% the result of the traveling shock crossing a standing re-collimation shock that is typically observed as the radio core
These events are illustrated in Fig. \ref{fig:lcurve} with vertical lines.

% ------
This study investigates the blazar 3C 454.3 using data from VLBA at radio frequencies of 43 GHz and 15 GHz. We analyze these observations alongside $\gamma$-ray flux variations in order to explore the jet structure, dynamics, emission mechanisms, and the location of $\gamma$-ray production regions.
Section \ref{sec:observations} details the observational data and the analysis methods employed. Section \ref{sec:results} presents our results and discusses the jet's physical properties. Finally, Sect. \ref{sec:summary} summarizes our key findings.
The cosmological parameters adopted throughout this paper are $H_{0}=71$ km $s^{-1} Mpc^{-1}$,   $\Omega_{\Lambda}=0.73$, and $\Omega_m=0.27$ \citep{2014A&A...571A..16P}.
At a redshift of z = 0.859, the distance to 3C 454.3 is estimated to be 5.489 Gpc. Given this distance, 1 mas on the sky corresponds to a physical size of 7.7 pc.
% ------

%--------------------------------------------------------------------
\section{Observational data and analysis} \label{sec:observations}
%--------------------------------------------------------------------
%%%%%%%%%%%%%%%%%%%%%%%%%%%%%%%%%%%%%
\begin{figure*}[ht!]
    \centering
    \includegraphics[width=0.9\textwidth]{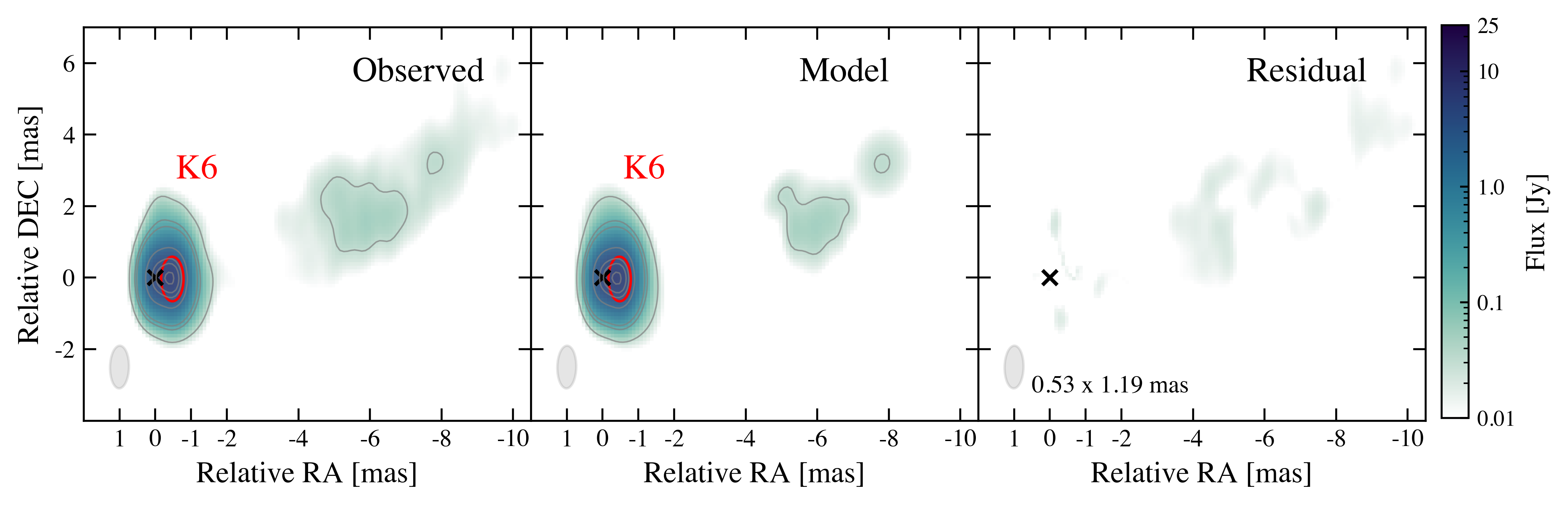}
    \caption{Left panel: VLBA intensity map at 15 GHz showing the position of K6 on November 21, 2020, convolved with a beam of 1.1 $\times$ 0.5 mas$^2$ at PA\,=\,1°.
    %VLBA intensity map from the 43 GHz observation of April 21, 2011, convolved with a beam of 0.33 $\times$ 0.14 mas$^2$ and PA\,=\,$-$10°. 
    Middle panel: Fitting result of the observed emission. Right panel: Residual image. The gray lines correspond to contour levels of 0.5, 3, 5, 30, 60, 75, and 95$\%$ of the peak total intensity.}
    \label{fig:mojave-map}
\end{figure*}
%%%%%%%%%%%%%%%%%%%%%%%%%%%%%%%%%%%%%%%%
%%%%%%%%%%%%%%%%%%%%%%%%%%%%%%%%%%%%%%%%
\begin{figure}[ht]
    \centering
    \includegraphics[width=0.7\columnwidth]{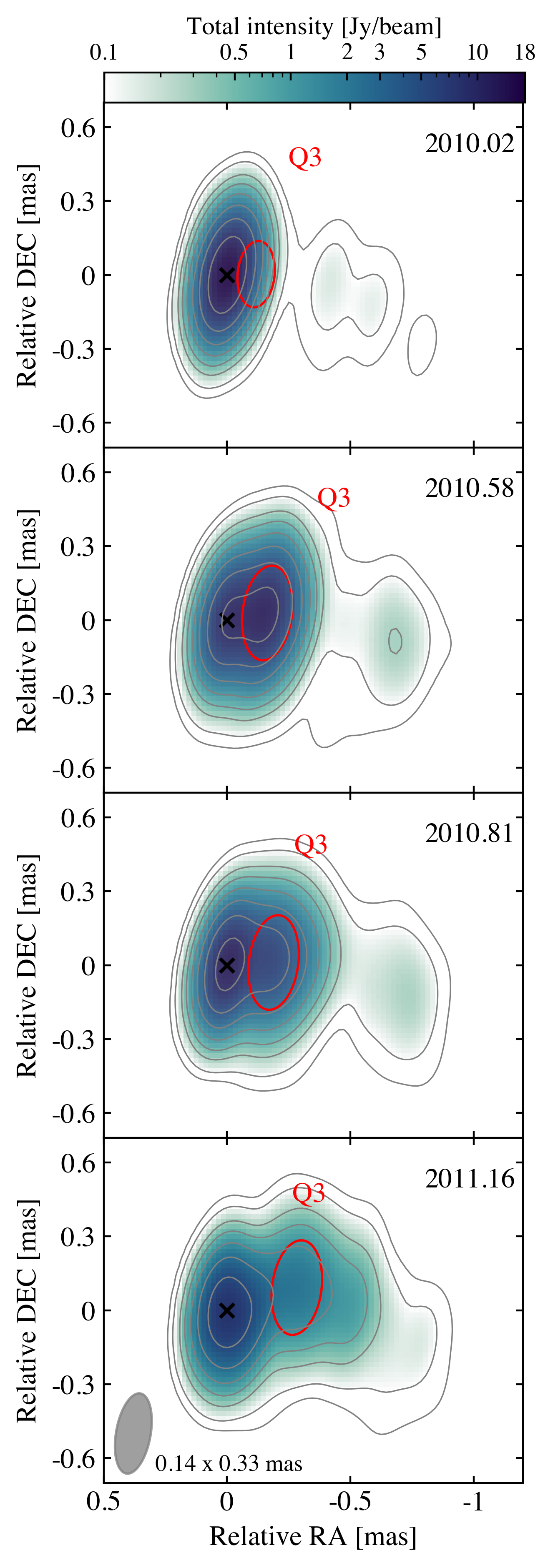}
    \caption{Sample sequence of the VLBA images at 43 GHz. Red ellipses across images show the position of the moving component Q3 at four different epochs.}
    \label{fig:q3}
\end{figure}
%%%%%%%%%%%%%%%%%%%%
This section describes the long-term radio and $\gamma$-ray datasets used in our analysis of 3C 454.3 as well as the methods we employed to analyze them. All the corresponding light curves are presented in Fig. \ref{fig:lcurve}.

\subsection{Gamma rays\label{gamma}}
% -----
We obtained $\gamma$-ray data for 3C 454.3 from the Fermi Large Area Telescope public archive, covering energies between 0.1 and 300 GeV \citep{2010ApJS..188..405A}. To process and analyze these data, we used the Fermitools software (version 2.0.8).
During the analysis, we considered all $\gamma$-ray sources within a 15-degree radius of 3C 454.3, as listed in the 4FGL-3DR catalog \citep{2022ApJS..260...53A}. We allowed the model to freely adjust the intensity of these nearby sources. For sources even closer (within a 5-degree radius), we allowed the model to adjust all their spectral parameters in addition to the intensity. This helped account for potential contamination from other $\gamma$-ray emitters in the region closer to our source of interest.
For our main analysis, we focused on $\gamma$-ray flux measurements centered on the specific dates when VLBA observations were made. Additionally, to illustrate the variability in the $\gamma$-ray emission, we created a weekly light curve spanning from August 2008 to late 2020. This light curve is shown in Fig. \ref{fig:lcurve}a, but it is important to note that this light curve is for demonstration purposes only and that it was not used in the analysis.
% -----

\subsection{VLBA observations at 43 and 15 GHz\label{vlba}}
We used 137 VLBA maps at 43 GHz (7 mm wavelength)  from the Boston University VLBA Blazar monitoring program (VLBA-BU-BLAZAR) and its successor, Blazars Entering the Astrophysical Multi-Messenger Era (BEAM-ME; \citealp{2017ApJ...846...98J}), to study the parsec-scale jet kinematics and structural evolution during the same period as the $\gamma$-ray light curve (2008 - 2020).\footnote{\url{http://www.bu.edu/blazars/BEAM-ME.html}} 
The program details, observations, and data reduction are described in \citet{2017ApJ...846...98J}. VLBA observations at 43 GHz typically achieve angular resolutions on the order of microarcseconds ($\mu$as) and sensitivities around a few tens of $\mu$Jy per beam. In the case of 3C 454.3, the angular resolution is on the order of $\sim$0.1 mas.

%\subsection{VLBA-MOJAVE 15 GHz observations\label{subsec:mojave}}
Additionally, we included 54 images at 15 GHz from the program Monitoring of Jets in Active Galactic nuclei with VLBA Experiments (MOJAVE; \citealt{2009AJ....137.3718L}).\footnote{\url{https://www.cv.nrao.edu/MOJAVE/project.html}} The images were taken from mid-2008 to late 2021. However, data from 2019 was excluded due to a known systematic effect causing a 10\% decrease in the flux density starting from the first half of 2019.\footnote{See also: \url{https://science.nrao.edu/enews/14.4/index.shtml\#vlba_flux}.} VLBA observations at 15 GHz can achieve sensitivities in the range of tens to hundreds of $\mu$Jy/beam and angular resolution on the order of tens to hundreds of $\mu$as. For 3C 454.3 the angular resolution varies depending on the number of available antennas and their configuration, but it has a typical value of $\sim$0.3 mas.

\subsection{OVRO 15 GHz observations\label{ovro}}
% ----
To complement the high-resolution VLBA data, we also examined the total flux density of 3C 454.3 at 15 GHz using data from the Owens Valley Radio Observatory (OVRO) blazar monitoring program (\citealp{2011ApJS..194...29R}).\footnote{\url{https://sites.astro.caltech.edu/ovroblazars/}} This data provides a broader view of the blazar emission compared to the VLBA observations, which focus on jet features. The light curve showing the variations in the total flux density at 15 GHz is included in Fig. \ref{fig:lcurve}d.
% ----
%\textbf{I might not be able to use these data given their data release policy... leaving this section pending...} 

\subsection{Sub-Millimeter Array 230 and 350 GHz \label{sma}}
% ---
To complement the radio data, we analyzed the blazar emission at higher frequencies of 230 and 350 GHz (wavelengths of 1.3 and 0.8 millimeters). The data was obtained from the publicly available Sub-Millimeter Array archive.\footnote{\url{http://sma1.sma.hawaii.edu/smaoc.html}} The light curve showing the variations in flux density at this frequency is included in Fig. \ref{fig:lcurve}b. For details on the observations and data processing, we refer the reader to \cite{2007ASPC..375..234G}.
% ---

%%%%%%%%%%%%%%%%%%%%%%%%%%%%%%%%%%%%%%%%%
%%%%%%%%%%%%%%%%%%%%%%%%%%%%%%%%%%%%%%%%%%%%%%%%%%%%%%%%%%%%%%%%%%%%%%%%%%%%%%%%%%%%%%%%%%
\begin{figure*}[ht!]
    \centering
    \includegraphics[width=\textwidth]{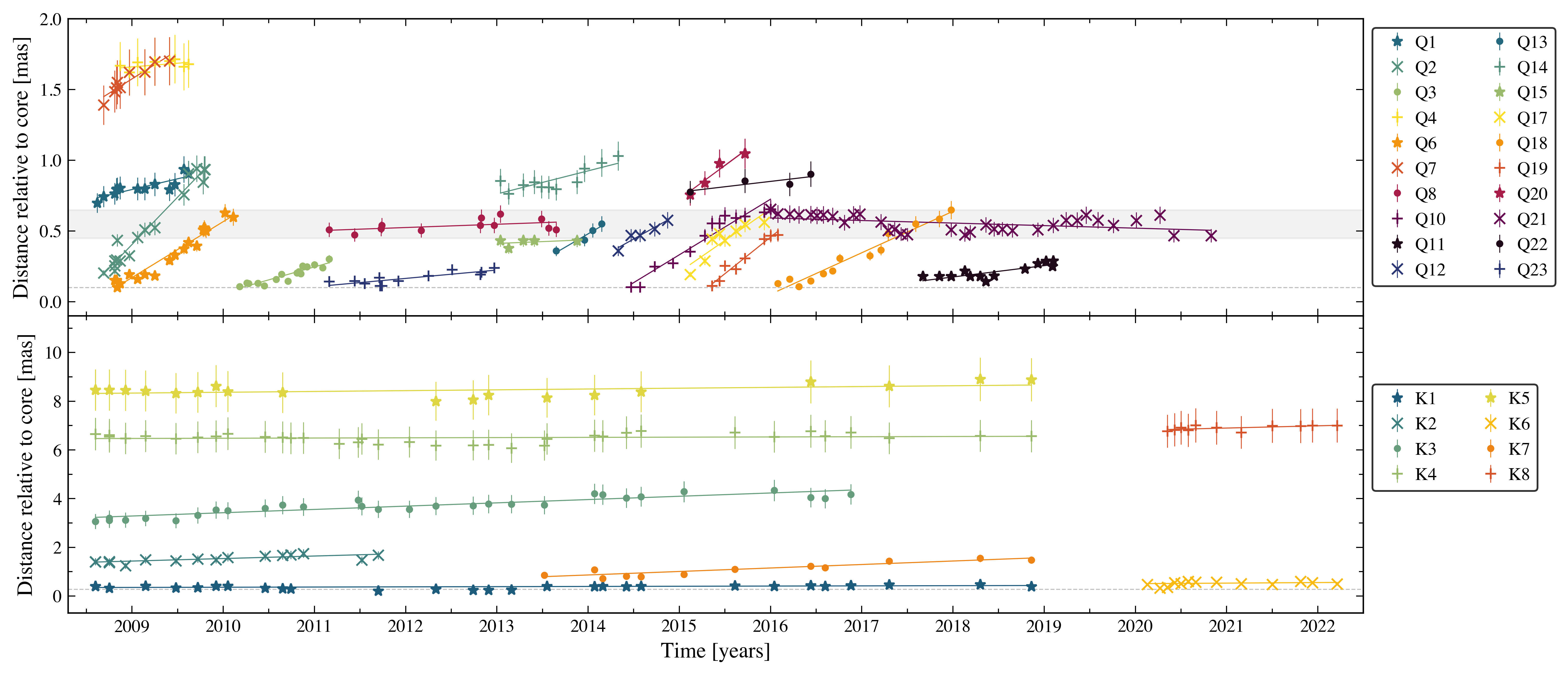}
    \caption{Relative distance to the core versus time of all the identified jet components in 3C 454.3 at 43 GHz (upper panel) and 15 GHz (lower panel). The gray shadowed area represents the position of the quasi-stationary region reported by this and previous studies. The dashed horizontal line shows the typical angular resolution.}
    \label{fig:dtocoreB}
\end{figure*}
%%%%%%%%%%%%%%%%%%%%%%%%%%%%%
% The dashed, horizontal gray line designates the position of the quasi-stationary region (ST) during our observing period, while 
%%%%%%%%%%%%%%%%%%%%%%%%%%%%%
%\begin{figure*}[ht!]
%    \centering
%    \includegraphics[width=0.8\textwidth]{knots-distance-core-mojave.png}
%    \caption{Relative distance to the core versus time of all the identified jet components in 3C 454.3 at 15 GHz.}
%    \label{fig:dtocoreM}
%\end{figure*}
%%%%%%%%%%%%%%%%%%%%%%%%%%%%%%%%%%%%%%%%%%%%%%%%%%%%%%%%%%%%%%%%%%%%%%%%%%%%%%%%%%%%%

% ------------------------------------------------------------------------
\subsection{Jet structure model fitting} \label{sec:modelfitting}
% ------------------------------------------------------------------------
To characterize the parsec-scale jet structure from the VLBA images, we followed the methodology described in \citet{2019A&A...630A..56P}.
We used the Common Astronomy Software Applications (CASA) task \texttt{IMFIT} to model the emission of each component, including the core, using 2D elliptical Gaussian components. 
% ----
A model feature was considered a detection if it equaled or exceeded 6$\sigma$ from the background emission. 
For both frequencies, we identified the core as the easternmost component along with other relevant knots in each map. Each component was cross-identified between different epochs by comparing the integrated flux and position relative to the core. 

The jet components, or knots, were named according to the frequency at which they were observed. Specifically, Q designates those observed at 43 GHz and K indicates those identified at 15 GHz. Knots were assigned sequential numbers according to the order in which they were observed within the jet. These designations also reflect the distance of the knot from the core, although this is not always the case. Components identified in fewer than five epochs were omitted (such as the components Q9 and Q16).

% ----
For comparison, we present the light curve of the total 43 GHz flux in Fig. \ref{fig:lcurve}, panel (c), alongside the core light curve at the same frequency.
The core light curve for the 15 GHz VLBA maps is shown in panel (d) of the same figure, along with the total flux at 15 GHz obtained from OVRO single-dish data.

\begin{figure*}[ht!]
    \centering
    \includegraphics[width=\textwidth]{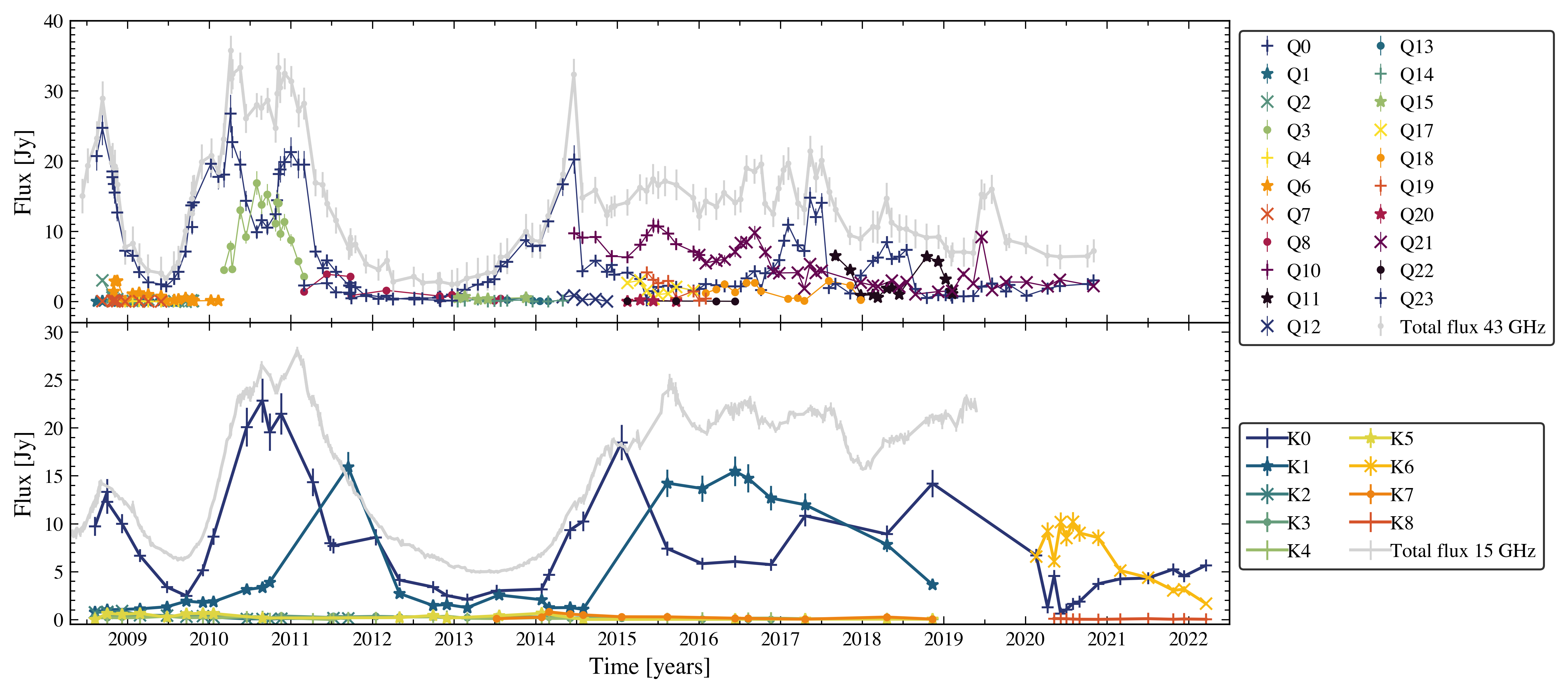}
    \caption{Light curves of all the identified jet components in 3C 454.3 at 43 and 15 GHz. }
    \label{fig:43lcurve}
\end{figure*}

%Mention Fig. \ref{fig:mojave-map} here. 
The observations made using high angular resolution VLBA revealed that 3C 454.3 can be categorized as core-dominated between 2008 and 2020, though with a few exceptional time intervals where the flux is dominated by some other component and displays an intricate multi-component jet structure.
Figure \ref{fig:mojave-map} illustrates the results of modeling using the task \texttt{IMFIT} near the core of 3C 454.3 within a margin of 2 and 10 mas at 43 and 15 GHz, respectively. 
We also tracked the temporal evolution of the radial separation between the different jet components and the radio core between 2008 and 2020 (see Fig. \ref{fig:q3}), enabling the calculation of their proper motions and apparent velocities (see Fig. \ref{fig:dtocoreB}). 
The light curves of all the identified jet components in 3C 454.3 at 43 and 15 GHz are shown in Fig. \ref{fig:43lcurve}.

%--------------------------------------------------------------------
\section{Results and discussion} \label{sec:results}
%--------------------------------------------------------------------
% ------------------------------------------------------------------

\begin{table}[!h]
\centering
\caption{Kinematic parameters of the identified jet components in the VLBA maps at 43 GHz.}
\setlength{\tabcolsep}{4pt} % Default value: 6pt
\renewcommand{\arraystretch}{1.2} % Default value: 1
\begin{tabular}{lcccrc}
    \hline
    ID  & $t_{eject}$ & $t_{cross}$ & $\mu$           & $\beta_{app}$ & d$_{core}$  \\
        & (yr)        & (yr)        & (mas yr$^{-1}$) & (c)           & (pc)        \\
    \hline
    \hline
%\textbf{P0} & -- & -- &-&-&0.0\\
Q1& 2003.8 & 2007.4 & 0.21$\pm$0.04 &9.8$\pm$1.9  & -- \\
Q2& 2008.4 & 2009.2 & 0.69$\pm$0.03 &32.2$\pm$1.6 & --\\
\rowcolor{lightgray!50}Q3&2009.7 &2012.7& 0.21$\pm$0.02 &9.9$\pm$1.1 & -- \\
Q4& 1977.0 & 1987.6& 0.29$\pm$0.23 & 13.6$\pm$11.0 & -- \\
%Q5 &  2015.6 & 1992.7& -- & -- & -- \\
Q6& 2008.6 & 2010.0& 0.42$\pm$0.02 & 19.5$\pm$1.1 &-- \\
Q7& 2005.1 & 2006.5& 0.46$\pm$0.07 & 21.7$\pm$3.1 &-- \\
Q8&  -- & -- & 0.02$\pm$0.06 & -- &  4.14$\pm$0.17  \\
%Q9 &  2009.2 & 2018.5& -- & -- & \\
Q10& 2014.1 & 2015.6&0.42$\pm$0.03 & 19.6$\pm$1.6 &--\\
Q11& 2015.8 & 2022.7&0.12$\pm$0.03 & 5.8$\pm$1.2 & --\\
Q12& 2013.3 & 2014.8&0.38$\pm$0.06 & 18.2$\pm$3.0 &--\\
Q13& 2012.7 & 2014.2&0.39$\pm$0.05 & 18.5$\pm$2.5 &--\\
Q14& 2008.3 & 2011.7&0.25$\pm$0.06 & 11.7$\pm$2.8 &--\\
Q15& 1997.3 & 2018.2&0.20$\pm$0.03 & 9.5$\pm$14.2 &--\\
%Q16 &  2043.6 & 2019.2&-- &-- & \\ % No tiene nada de sentido
Q17& 2014.5 & 2015.8&0.55$\pm$0.01 & 25.9$\pm$5.2 & --\\
Q18& 2015.4 & 2016.8&0.63$\pm$0.52 & 29.9$\pm$24.5&--\\
Q19& 2015.1 & 2014.6&0.59 $\pm$0.08  & 27.6$\pm$3.7 &--\\
Q20& 2013.5 & 2014.6&0.52$\pm$0.07 & 24.2$\pm$3.6 &--\\
\rowcolor{lightgray!50}Q21&  -- & --& -0.02$\pm$0.03 &-- & 4.28$\pm$0.13\\
Q22& 2009.7 & 2013.6&0.25$\pm$0.11 & 11.9$\pm$5.4 &--\\
Q23& 2009.2 & 2018.5&0.08$\pm$0.02 & 3.9$\pm$0.8 &-- \\
Q24& 2015.8 & 2017.7&0.3$\pm$0.02  & 14.3$\pm$0.9&-- \\
\hline
\end{tabular}
\tablefoot{Columns from left to right: (1) component ID, (2) estimation of the time in which it was ejected from the core,  (3) time at which it crosses the quasi-stationary region, (4) proper motions, (5) apparent speed, and (6) projected distance to the core in pc. Q8 and Q21 are the quasi-stationary components identified between 2011-2013 and 2016-2020, respectively.}
\label{tab:kinematicB}
\end{table}

%--------------------------------------------------------------------

% ------------------------------------------------------------------------
\subsection{Jet structure and kinematics} \label{sec:structure&kinematics}
% ------------------------------------------------------------------------
%describir la evolución estructural del jet a 43 GHz
In our analysis of the VLBA maps at 43 GHz, we characterized the core (Q0), which we assumed to be a stationary element situated at the eastern end of the jet, along with several other notable features (see Fig. \ref{fig:q3}). 
 % estacionaria 
We observed a quasi-stationary component oscillating between 0.5 and 0.6 mas, which was visible during two different time intervals: Q8 between 2011 and 2014 and Q21 from 2016 to 2020. 
This component was previously observed and labeled as St (e.g., \citealp{2005AJ....130.1418J, 2013ApJ...773..147J}) or region C (e.g., \citealp{2017ApJ...846...98J, 2024A&A...682A.154T}).
 % eyecciones moviles
Additionally, we identified 20 moving components, labeled as Q1, Q2, and so on up to Q24 (excluding Q8 and Q21, as already mentioned). Most of the modeled components are consistent with knots discussed in earlier studies.
Figure \ref{fig:q3} presents four different epochs where we identified component Q3, ejected from the core in early 2010. This ejection coincides with the $\gamma-$ray flare observed between September 2009 and January 2010. Previously, \citet{2013ApJ...773..147J} identified this component as K09, and its passage through the core was associated with a multi-wavelength outburst in 3C 454.3 \citep{2013ApJ...763L..36L, 2013ApJ...773..147J, 2021ApJ...906....5A}.

\begin{table}[h!]
\centering
\caption{Kinematic parameters of the identified jet components in the VLBA maps at 15 GHz.}
\setlength{\tabcolsep}{8pt} % Default value: 6pt
\renewcommand{\arraystretch}{1.2} % Default value: 1
\begin{tabular}{lcccc}
    \hline
    ID  & $t_{eject}$ & $\mu$            &$\beta_{app}$ & d$_{core}$ \\
        &  (yr)        & (mas yr$^{-1}$) & (c)          & (pc)       \\
    \hline
    \hline
K1          & --     & 0.01$\pm$0.02  & --        & 2.83$\pm$0.26  \\
K2          & 1995.1 & 0.17$\pm$0.04 & 8.0$\pm$1.7& -- \\
K3          & 1984.7 & 0.18$\pm$0.01 & 7.7$\pm$0.6& -- \\
K4          &  --    &0.01$\pm$0.02  &--         & 50.0$\pm$0.03 \\
K5          &  --    &0.03$\pm$0.03  &--         & 64.9$\pm$0.05 \\
\rowcolor{lightgray!50}K6 & 1997.8&0.22$\pm$0.25& -- & 3.9$\pm$0.1\\
K7          & 2007.9    &0.18$\pm$0.02  &8.3$\pm$1.2       & -- \\
\rowcolor{lightgray!50}K8 & -- &0.09$\pm$0.09& -- & 53.18$\pm$0.01\\
\hline
\end{tabular}
\tablefoot{Columns from left to right: (1) Component ID, (2) estimation of the time in which it was ejected from the core, (3) proper motions, (4) apparent speed, and (5) projected distance to the core in parsecs.}
\label{tab:kinematicM}
\end{table}

% los que desaparecen de Efthalia
Similarly, Q10, Q17, Q19, and Q18 seem to align with the knots referenced by \citet{2024A&A...682A.154T} as the disappearing knots observed at 43 and 86 GHz. These four components ejected from the core in February 2014, July 2014, February 2015, and May 2015, respectively, seemed to vanish once they approached the quasi-stationary component Q21 (or region C in their study), as depicted in Fig. \ref{fig:dtocoreB}. \citet{2024A&A...682A.154T} proposed that this disappearance might be attributed to the bending of the jet at that specific location, thus causing the plasma to shift toward the observer's line of sight.
%Faltan Weaver y Liodalkis

%qué mas puedo decir del core? 
At this particular frequency, the emission is mainly dominated by Q0, with only occasional periods when this is not the case. 
For instance, in August 2010, Q3 reached a maximum flux of 16.9 Jy/beam (see Fig. \ref{fig:q3}), which corresponds to just over 170$\%$ of the core flux at the same time.
Between 2014 and 2016, the flux was primarily dominated by the mobile component Q10, which surpassed the core flux by a factor of seven. In 2016, the emission was primarily dominated by the quasi-stationary component Q21 and had an average flux twice that of the core for that year.
 
% 15 GHz
%describir la evolución estructural del jet a  15 GHz 
Figure \ref{fig:mojave-map} presents the results of model fitting applied to 3C 454.3 at 15 GHz. The image reveals a complex jet structure oriented toward the northwest that exhibits subtle variations over time. This finding is further supported by observations from August 2008 to the end of 2020, which consistently depict the source as core dominated with a persistent one-sided jet.

%estacionarias
The brightest component, designated as K0, is followed by K1, a quasi-stationary jet feature located on average 0.4 mas away. Notably, K1 remained attached to the core from mid-2008 to 2013. By the end of 2020, a new component, K6, emerged. K6 originated from the initial position of K1 and moved away from K0, and it is potentially linked to the $\gamma$-ray emission, as we discuss later.
Given that we excluded the 15 GHz VLBA 2019 data from our analysis (due to the reported flux discrepancies), we cannot definitively confirm if K6 originated from the same K1 component observed earlier. In addition to the moving components and the core, we identified two other quasi-stationary features, K4 and K5, located at approximately 6.5 mas and 8.5 mas from the core, respectively.

% apparent speed
We calculated the apparent speed ($\beta_{app}$) of the moving components following the relationship
\begin{equation}
    \beta_{app} = \frac{\mu D_L}{c(1+z)},
\end{equation}
where $\mu$ is the proper motion in radians per second, $D_L$ is the luminosity distance in meters, $z$ is the source redshift, and $c$ is the speed of light in meters per second. 
%%%%%%%%%%%%%%%%%%%%%%%%%%%%%
%%%%%%%%%%%%%%%%%%%%%%%%%%%%%
% ------------------------------------------------------------------
\begin{table}[!h]
\centering
\caption{Spearman rank correlation coefficients obtained between the component fluxes in the VLBA maps at 43 GHz and the simultaneous $\gamma$-ray fluxes.}
\setlength{\tabcolsep}{8pt} % Default value: 6pt
\renewcommand{\arraystretch}{1.4} % Default value: 1
\begin{tabular}[]{lrrrr}
    \hline
    ID  & $\rho$ & $\rho^2$ & p-value & $\rho_z$    \\
    \hline
    \hline
\rowcolor{lightgray!50}Q0 &   0.52 &   0.27 & $2.4\times10^{-9}$ & $0.519_{-0.012}^{+0.012}$ \\
Q1 &  $-0.33$ &   0.11 &      0.27 & $-0.26_{-0.091}^{+0.096}$ \\
Q2 &   0.10 &   0.01 &      0.70  & $0.123_{-0.057}^{+0.056}$ \\
\rowcolor{lightgray!50}Q3 &   0.53 &   0.28 & 0.01  & $0.510_{-0.024}^{+0.024}$ \\
Q4 &  $-0.42$ &   0.17 &      0.26  & $-0.483_{-0.086}^{+0.096}$ \\
Q5 &  $-0.54$ &   0.29 &      0.27  & $-0.543_{-0.053}^{+0.058}$ \\
Q6 &  $-0.30$ &   0.09 &      0.21  & $-0.306_{-0.054}^{+0.056}$ \\
Q7 &   0.31   &   0.10 &      0.35  & $0.264_{-0.099}^{+0.094}$ \\
Q8 &   0.05   &   0.00 &      0.90  & $0.150_{-0.178}^{+0.169}$ \\
Q9 &   0.20   &   0.04 &      0.70  & $0.200_{-0.130}^{+0.124}$ \\
Q10 &   0.45   &   0.20 &      0.14  & $0.420_{-0.108}^{+0.098}$ \\
Q11 &  $-0.61$ &   0.37 &      0.15  & $-0.464_{-0.138}^{+0.165}$ \\
Q12 &  $-0.15$ &   0.02 &      0.65  & $-0.098_{-0.089}^{+0.091}$ \\
Q13 &  $-0.50$ &   0.25 &      0.39  & $-0.500_{-0.178}^{+0.234}$ \\
Q14 &  $-0.49$ &   0.24 &      0.15  & $-0467_{-0.086}^{+0.096}$ \\
Q15 &   0.07   &   0.01 &      0.88  & $0.107_{-0.141}^{+0.137}$ \\
Q16 &  $-0.38$ &   0.15 &      0.28  & $-0.297_{-0.117}^{+0.127}$ \\
Q17 &  $-0.32$ &   0.10 &      0.48  & $-0.321_{-0.097}^{+0.105}$ \\
Q18 &   0.50   &   0.25 &      0.67  & $-0.396_{-0.083}^{+0.090}$ \\
Q19 &   0.30   &   0.09 &      0.62  & $0.300_{-0.054}^{+0.052}$ \\
%Q20 &  $-0.80$ &   0.64 &      0.20  & $_{-}^{+}$ \\
\rowcolor{lightgray!50}Q21 &   0.45  &   0.20 & 0.01  & $0.444_{-0.064}^{+0.060}$ \\
Q22 &   0.52   &   0.27 &      0.29  & $0.543_{-0.270}^{+0.190}$ \\
Q23 &   0.20   &   0.04 &      0.70  & $0.200_{-0.132}^{+0.125}$ \\
%Q24 &  $-0.44$ &   0.19 &      0.13  & $_{-}^{+}$ \\
\hline
\end{tabular}
\tablefoot{Columns from left to right: (1) Component ID, (2) Spearman rank correlation coefficient, (3) correlation coefficient squared, (4) p-value (Q0 corresponds to the easternmost component identified as the 43 GHz core), (5) correlation coefficient calculated using the z-transform, $\rho_z$, and errors. Q21 is the quasi-stationary component identified between 2016 and 2020. The gray shaded lines indicate components with a p-value of less than 5$\%$.}
\label{tab:spearmanB}
\end{table}

%--------------------------------------------------------------------
\subsection{Correlation analysis} \label{sec:correlation}
%--------------------------------------------------------------------
The connection between the $\gamma$-ray state of blazars and their parsec-scale jet behavior is well established. Studies by \citealp{2011A&A...532A.146L} suggested that the brightest $\gamma$-ray flares in blazars might be triggered by the initial stages of millimeter-wavelength outbursts. 
To quantify the potential link between the variability of the parsec-scale jet knots in 3C 454.3 and its $\gamma$-ray emission, we performed Spearman’s rank correlation tests. This non-parametric test assesses the strength and direction of a monotonic relationship between two variables, independent of their underlying distributions.
%between the simultaneous flux densities of each knot and their $\gamma$-ray fluxes. 
%We calculated $\gamma$-ray fluxes in time bins centered around the observation times of the VLBA data. 
To account for the possibility of rapid $\gamma$-ray variations reported in previous studies (e.g., \citealt{2016ApJ...824L..20A, 2018ApJ...856...99P}), we implemented an adaptive time binning approach. We calculated the $\gamma$-ray fluxes within time bins centered on the exact observation times of the VLBA data. The bin sizes were iteratively increased, starting small and always maintaining the VLBA observation date as the center. This process continued until the test statistic for each bin exceeded a threshold of 25. We consider a given correlation as significant if the probability of obtaining a specific correlation coefficient due to statistical fluctuations is less than 5$\%$ (i.e., p-value $< 0.05$). This approach ensured we captured potential correlations while maintaining the temporal resolution necessary to detect rapid variability.
% To test correlations

To further test the robustness of the observed correlations and minimize potential biases from individual data points or small subsets, we performed a jackknife analysis on each component. This analysis, which involves iteratively removing one data point and recalculating the correlation, demonstrated consistent results. The mean correlation coefficients and p-values obtained from the jackknife analysis were in good agreement with those presented in Tables \ref{tab:spearmanB} and \ref{tab:spearmanM} in all cases.
Also, to assess the reliability of the presented correlations, we adopted the methodology described in \cite{1997ASSL..218..163A}. We conducted 10,000 Monte Carlo simulations, generating simulated datasets with fluxes drawn randomly from within their respective error ranges.
Spearman rank correlations were then computed for each simulated dataset.
Subsequently, we performed Fisher's z-transform on the resulting correlation coefficients, calculating the median and standard deviation in the z-space.
The median was then back-transformed to obtain a correlation coefficient, and asymmetric errors were estimated.
The last column of Tables \ref{tab:spearmanB} and \ref{tab:spearmanM} present the correlation coefficients and associated errors determined through this Monte Carlo approach, and they are consistent with the Spearman correlations obtained from the original data.

\begin{figure}[h!]
    \centering
    \includegraphics[scale=0.65]{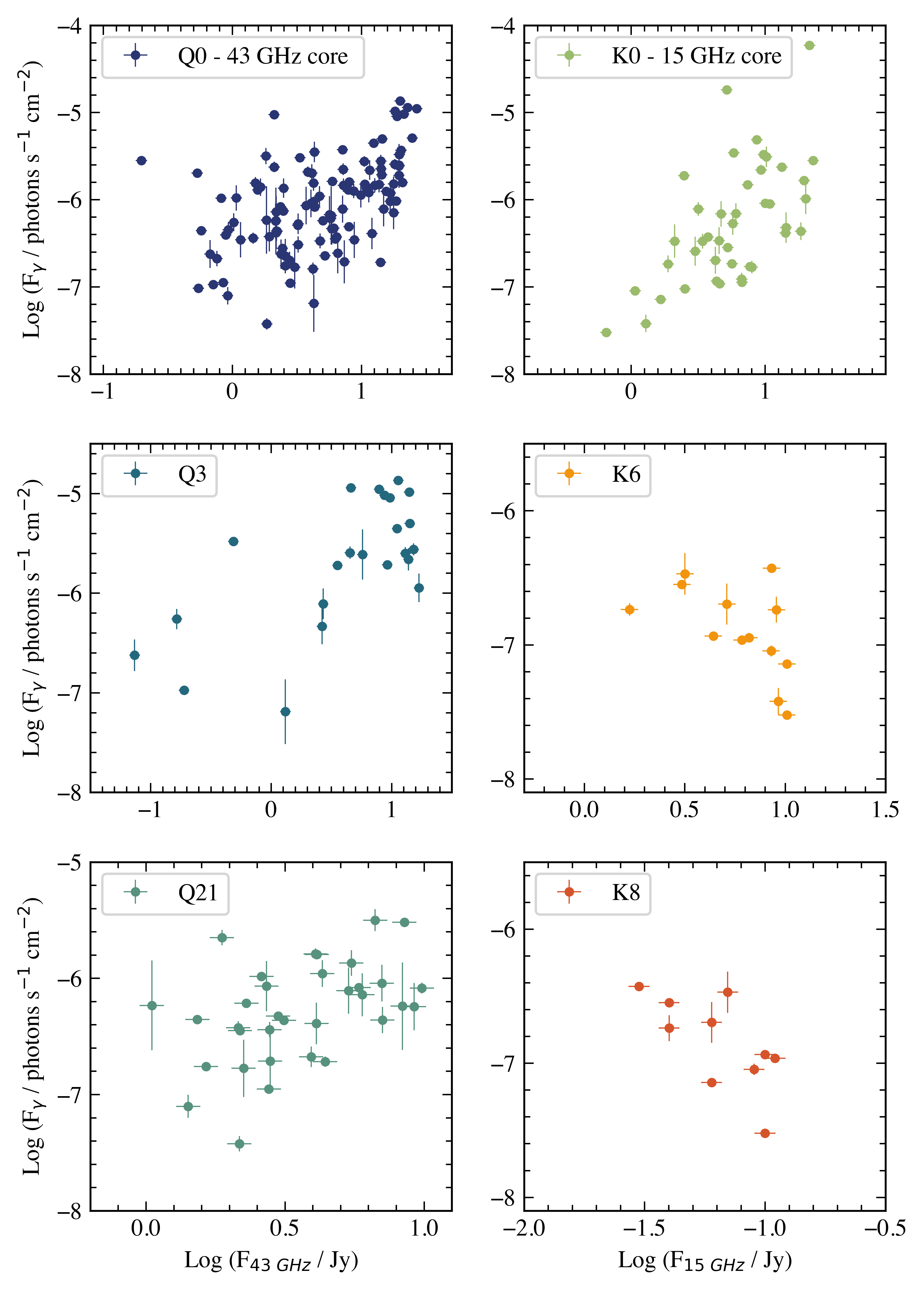}
    \caption{Plots of $\gamma-$flux versus VLBI-flux for the components with significant correlation. Left: $\gamma$-ray flux versus 43 GHz flux for the Q0 (core), Q3 and Q21 
    components in the VLBA maps. Right: $\gamma$-ray flux versus 15 GHz flux for the K0 (core), K6 and K8 components.} 
    \label{fig:fluxflux}
\end{figure}

% correlation at 43 GHz
The results of our analysis are presented in Table \ref{tab:spearmanB}, they reveal significant correlations between the $\gamma$-ray emission and the 43 GHz flux densities of specific jet components found in the BEAM-ME maps. The core (Q0) shows the strongest correlation, with a coefficient ($\rho$) of 0.52. 
This suggests a possible link between core variability at 43 GHz and roughly 27$\%$ of the observed $\gamma$-ray variability, as indicated by a Spearman rank correlation coefficient squared ($\rho^2$) of 0.27.
Similarly, correlations with components Q3 and Q21 suggest that 26$\%$ and 18$\%$ of the $\gamma$-ray variability, respectively, could be related to the 43 GHz emission in these components during their detection periods.
Figure \ref{fig:fluxflux} illustrates these relationships between the simultaneous $\gamma$-ray fluxes and the 43 GHz flux densities of the core and components Q3 and Q21.
The data points associated with Q8 and Q21 correspond to two distinct time periods when the quasi-stationary component was observed. This component has been extensively studied in the literature prior to this work (e.g., \citealt{1987Natur.328..778P}, \citealt{2001ApJ...556..738J}) and before the Fermi era. Intriguingly, during the first epoch, Q8 exhibits no correlation with the gamma-ray variability, while Q21 shows a clear correlation. As illustrated in Fig. \ref{fig:dtocoreB}, numerous moving features intersect with Q21, whereas none pass through Q8. This observation aligns with models proposed in previous studies that attribute $\gamma$-ray flares to shock-shock interactions between moving and stationary features (e.g., \citealt{1985ApJ...298..114M, 2011A&A...531A..95F}). These findings suggest that the quasi-stationary component may act as a $\gamma$-ray emission region, with its activity being modulated by the occurrence of such interactions.

% ---- correlations at 15 GHz - >> core + anti-correlations
Figure \ref{fig:fluxflux} (right panels) also depicts the relationships between the $\gamma$-ray fluxes and the 15 GHz flux densities of the core (K0) and components K6 and K8 from the MOJAVE maps. Table \ref{tab:spearmanM} summarizes the Spearman rank correlation results.
The core, once again, exhibits a strong and significant correlation with $\gamma$-ray emission, with a coefficient of 0.62. This implies that roughly 39$\%$ of the $\gamma$-ray variations are likely linked to the core 15 GHz activity.
Interestingly, a new jet component, K6, emerged near the end of the study period (early 2020). This component shows a significant anti-correlation with the $\gamma$-ray flux, with a coefficient of $-0.65$, suggesting that nearly 40$\%$ of the $\gamma$-ray variability during this specific time might be related to the ejection and 15 GHz emission of K6, even in the absence of a simultaneous flare. 
As suggested by \citet{2018MNRAS.479.2037P}, one possible explanation for the observed behavior is that the jet became less transparent to $\gamma$-rays. This change in transparency might be caused by gamma-ray absorption due to triple pair production.
%the ejection of new material into the jet, a process often associated with flares. 
%This new material could accelerate the $\gamma$-ray emitting region, making it more likely for high-energy particles to interact with the $\gamma$-rays, triggering triple pair production. 
The cross section for both the IC scattering and the triple pair production are dependent on the energy of the electrons in the jet, as when the Lorentz factors of individual electrons increase up to a certain threshold, the triple pair production becomes more prominent than the IC.
This process would absorb $\gamma$-rays, causing a flux decrease, while at the same time increasing the synchrotron emission, due to the increase in the Lorentz factor, and the production of more leptons within the jet.
However, the proximity of K6 to the core as seen in Fig. \ref{fig:mojave-map}, near the spatial resolution limit, raises the possibility of flux contamination.
Additionally, a quasi-stationary component, K8, was detected at approximately 53 pc from the core. Although correlation analysis was feasible with 11 epochs, more observations are essential to determine this relationship.

%% -----

This study offers novel insights into the $\gamma$-ray emission in blazars. Our findings reveal the presence of multiple distinct regions within the jet, with each contributing to the high-energy emission. 
Notably, some of these regions exhibit non-stationary behavior and display temporal variability. Furthermore, we successfully located these emission regions within the jet structure.
Previous multi-frequency studies of 3C 454.3 have revealed strong correlations between $\gamma$-ray and optical emissions (\citealt{2011ApJS..195...19O, 2013ApJ...763L..36L, 2016Galax...5....3V, 2019A&A...631A...4N, 2021ApJ...906....5A, 2021ApJS..253...10F}). 
These correlations strongly suggest that $\gamma$ rays in blazars originate from leptonic processes rather than hadronic ones, as no known hadronic mechanism can simultaneously generate $\gamma$ rays and significantly impact optical synchrotron emission.
To distinguish between SSC and EIC as the dominant leptonic process, determining the location of the emitting region within the jet is crucial. For regions located several parsecs away from the central engine, the density of external seed photons is typically lower than the synchrotron photon density, thus favoring SSC as the primary emission mechanism.
This is supported by findings in 3C 279, where \citet{2019A&A...630A..56P} concluded that approximately 33\% of the observed $\gamma$-ray emission originates from SSC processes within a particular jet component during its the observable period.
In the core region of the blazar, while we can confirm a leptonic origin for the gamma-ray emission, differentiating between SSC and EIC remains challenging. However, it is generally accepted that sufficiently energetic electrons exist in the core to power the observed $\gamma$-ray emission.
For non-core components (i.e., those located farther from the central parsec), the correlation observed in simultaneous data suggests that the $\gamma$-ray emission mechanism is SSC. 
The electron energy distribution within a knot is such that while lower-energy electrons dominate the radio band, higher-energy electrons capable of producing gamma-rays can co-exist.
These findings can guide the development of more comprehensive theoretical models that might account for the observed variability and spatial distribution of high-energy emissions.
%% ---

% ----
%The respective relations between the fluxes of the core and the component E6 in the MOJAVE 15 GHz maps with the simultaneous $\gamma$-ray fluxes are shown in Fig. \ref{fig:fluxflux}. The results of the Spearman rank correlation tests are summarized in Table \ref{tab:spearmanM}: 
% correlation results at 15 GHz
%The core variability at 15 GHz shows a strong and significant correlation with the $\gamma$-ray emission, displaying a correlation coefficient of 0.55 this indicates that roughly 30$\%$ of the fluctuations occurring in $\gamma$ rays can be attributed to, or linked with, the variations in the core flux density at 15 GHz. 
%Interestingly, towards the end of the study period, in early 2020, we detected the appearance of a new jet component, labeled E6, which displayed a strong and significant anti-correlation with the $\gamma$-ray flux with a coefficient of -0.89, which implies that close to 80$\%$ of the $\gamma$-ray emission could be related to the ejection and 15 GHz emission of this particular feature during a time where no flare behavior is observed simultaneously.
%discussion
%This is important, because not only is it being shown that there is more than one region of $\gamma$-ray emission, but also that in certain periods of time, some of these regions are not stationary. In addition to that, the exact location of these emission regions in the blazar has been identified. These are important parameters, both for models and for understanding the complex nature of these sources.
\begin{table}[h!]
\centering
\caption{Spearman rank correlation coefficients obtained between the component fluxes in the VLBA maps at 15 GHz and the simultaneous $\gamma$-ray fluxes.}
\setlength{\tabcolsep}{8pt} % Default value: 6pt
\renewcommand{\arraystretch}{1.4} % Default value: 1
\begin{tabular}{lrrrr}
    \hline
    ID  & $\rho$ & $\rho^2$ & p-value & $\rho_z$\\
    \hline
    \hline
\rowcolor{lightgray!50}K0 &0.62 &  0.39 &  $1\times10^{-5}$ & $0.619_{-0.016}^{+0.016}$ \\
K1 &   0.06 &   0.00 &     0.84  & $0.071_{-0.029}^{+0.029}$ \\
K2 &  -0.07 &   0.01 &     0.80  & $-0.046_{-0.031}^{+0.031}$ \\
K3 &   0.08 &   0.01 &     0.71  & $0.099_{-0.042}^{+0.042}$ \\
K4 &   0.22 &   0.05 &     0.26  & $0.208_{-0.04}^{+0.04}$ \\
K5 &  -0.07 &   0.00 &     0.82  & $-0.064_{-0.077}^{+0.078}$ \\
\rowcolor{lightgray!50}K6        &  -0.65 &   0.42 & 0.02  & $-0.632_{-0.062}^{+0.071}$ \\
K7 &   0.35 &   0.12 &     0.30  & $0.382_{-0.078}^{+0.073}$ \\
\rowcolor{lightgray!50}K8 &  -0.64 &   0.40 & 0.04  & $-0.636_{-0.059}^{+0.068}$ \\
\hline
\end{tabular}
\tablefoot{Columns are the same as in Table \ref{tab:spearmanB}. K0 corresponds to the 15 GHz core. Gray shaded lines indicate components with significant correlations.}
\label{tab:spearmanM}
\end{table}
%%%%%%%%%%%%%%%%%%%%%%%%%%%%%%%%
%%%%%%%%%%%%%%%%%%%%%%%%%%%%%%%%

%--------------------------------------------------------------------
\section{Summary and conclusions} \label{sec:summary}
%--------------------------------------------------------------------
In this study, we have analyzed the jet behavior and structure of the blazar 3C 454.3 using radio data from the VLBA-BU-BLAZAR, BEAM-ME, and MOJAVE monitoring programs (at frequencies of 43 GHz and 15 GHz). We investigated potential $\gamma$-ray emission regions within the jet components identified in the VLBA maps. This included measuring the distances between these regions and the radio core and exploring possible emission mechanisms. Additionally, we examined whether these regions account for all the $\gamma$-ray emission or, as in the case of 3C 279 \citep{2019A&A...630A..56P}, multiple regions are needed to describe it. Our key findings include the following:

\begin{itemize}
    \item Both 43 GHz and 15 GHz core emissions strongly correlate with $\gamma$-ray emission. We observed a robust correlation between the 43 GHz core emission and $\gamma$-ray emission, with a correlation coefficient of 0.52. This indicates that approximately 27$\%$ of the variability in the $\gamma$-ray emission is related to the variability in the core radio emission. Similarly, the flux of the 15 GHz core exhibits a strong and statistically significant correlation with the $\gamma$-ray emission, contributing approximately 40$\%$ to the high-energy emission.
    
    \item The quasi-stationary component Q21 in the 43 GHz maps, visible between 2016 and 2021, at a projected distance of 4.6 pc (0.6 mas) strongly correlates with the $\gamma$-ray flux, implying that it may contribute up to 18\% of the high-energy emission. However, the quasi-stationary component observed during a different time interval, Q8, does not correlate. This variability in the correlation suggests that quasi-stationary components may exhibit intermittent gamma-ray emission.
    
    \item At 43 GHz, we detected a mobile component whose flux exhibits a robust correlation with the $\gamma$-ray flux, contributing 26$\%$ of the high-energy emission between 2010.18 and 2011.16. This component, denoted as Q3, demonstrates an apparent velocity of $\beta_{app}=9.9\pm1.1$ c, observed at a projected distance ranging from 0.8 to 2.3 pc (0.1 to 0.3 mas).
 
    \item For $\gamma$-ray emission regions located farther than one parsec from the central black hole (e.g., components Q21 and Q3), our findings strongly suggest SSC as the primary mechanism responsible for the high-energy radiation. This conclusion is supported by the observed correlation between the radio variability in these regions and the $\gamma$-ray emission, indicating a nearly simultaneous variability.
    
   \item In the 15 GHz maps, a new jet component, K6, appeared in early 2020. It strongly anti-correlates with $\gamma$-ray flux (correlation coefficient -0.65), suggesting a potential link to around 40$\%$ of the $\gamma$-ray variability. 
   However, the proximity of K6 to the core might distort its measured flux. Another component, K8, shows a similar anti-correlation with $\gamma$-rays, but we believe more data is needed to confirm this behavior.

   \item The observed anti-correlations in the two 15 GHz components may be attributed to an increased Lorentz factor in the electrons of the $\gamma$-ray emission region, which can enhance electron-positron triple pair production and lead to greater $\gamma$-ray absorption and a subsequent decrease in observed gamma-ray flux, as suggested by \citet{2018MNRAS.479.2037P}. This would also increase the synchrotron emission because the energy and number of the leptons are increasing, thus creating the observed anti-correlation.
  
\end{itemize}

Our findings are valuable on two fronts. First, they provide crucial data for refining theoretical models of blazar jets. Second, they open new avenues for investigation by pinpointing specific regions within the jet for further study. This comprehensive understanding might lead to a deeper grasp of the complex processes at play in blazars.
%%%%%%%%%%%%%%%%%%%%
\begin{acknowledgements}
    EP would like to express her sincere gratitude to the faculty and graduate students of the Max Planck Institute for Radio Astronomy (MPIfR) for the insightful discussions and generous hospitality extended during her visit. 
    Particular thanks are extended to Dr. D. Kim for his dedication of time and for engaging in insightful discussions concerning the analysis of $\gamma$-ray data. 
    We thank Dr. G. Paraschos for his role as the internal referee at the MPIfR.
    This work is supported by CONAHCyT research grants 280789 and 320987, and by the MPIfR-Mexico Max Planck Partner Group led by VMP-A.
    This publication is part of the M2FINDERS project which has received funding from the European Research Council (ERC) under the European Union’s Horizon 2020 Research and Innovation Programme (grant agreement No. 101018682).
    This research has made use of data from the OVRO 40 m monitoring program \citep{2011ApJS..194...29R}, which is supported in part by NASA grants NNX08AW31G, NNX11A043G, and NNX14AQ89G and NSF grants AST-0808050 and AST-1109911. 
    This research has used data from the MOJAVE database, which is maintained by the MOJAVE team \citep{2009AJ....137.3718L, 2018ApJS..234...12L}.  
    This study makes use of VLBA data from the BEAM-ME and VLBA-BU-BLAZAR Monitoring Programs ( \url{http://www.bu.edu/blazars/BEAM-ME.html}), funded by NASA through the Fermi Guest Investigator Program. 
    The VLBA is an instrument of the National Radio Astronomy Observatory. The National Radio Astronomy Observatory is a facility of the National Science Foundation operated by Associated Universities, Inc.
    The Submillimeter Array is a joint project between the Smithsonian Astrophysical Observatory and the Academia Sinica Institute of Astronomy and Astrophysics and is funded by the Smithsonian Institution and the Academia Sinica. We recognize that Maunakea is a culturally important site for the indigenous Hawaiian people; we are privileged to study the cosmos from its summit.

\textit{Software.} Astropy \citep{2013A&A...558A..33A,2018AJ....156..123A}, CASA \citep{2007ASPC..376..127M}, Fermitools v 2.0.8 (\url{https://fermi.gsfc.nasa.gov/ssc/data/analysis/}).

\end{acknowledgements}
 
\bibliography{aa52255-24}{}

@ARTICLE{2018AJ....156..123A,
       author = {{Astropy Collaboration} and {Price-Whelan}, A.~M. and {Sip{\H{o}}cz}, B.~M. and {G{\"u}nther}, H.~M. and {Lim}, P.~L. and {Crawford}, S.~M. and {Conseil}, S. and {Shupe}, D.~L. and {Craig}, M.~W. and {Dencheva}, N. and {Ginsburg}, A. and {VanderPlas}, J.~T. and {Bradley}, L.~D. and {P{\'e}rez-Su{\'a}rez}, D. and {de Val-Borro}, M. and {Aldcroft}, T.~L. and {Cruz}, K.~L. and {Robitaille}, T.~P. and {Tollerud}, E.~J. and {Ardelean}, C. and {Babej}, T. and {Bach}, Y.~P. and {Bachetti}, M. and {Bakanov}, A.~V. and {Bamford}, S.~P. and {Barentsen}, G. and {Barmby}, P. and {Baumbach}, A. and {Berry}, K.~L. and {Biscani}, F. and {Boquien}, M. and {Bostroem}, K.~A. and {Bouma}, L.~G. and {Brammer}, G.~B. and {Bray}, E.~M. and {Breytenbach}, H. and {Buddelmeijer}, H. and {Burke}, D.~J. and {Calderone}, G. and {Cano Rodr{\'\i}guez}, J.~L. and {Cara}, M. and {Cardoso}, J.~V.~M. and {Cheedella}, S. and {Copin}, Y. and {Corrales}, L. and {Crichton}, D. and {D'Avella}, D. and {Deil}, C. and {Depagne}, {\'E}. and {Dietrich}, J.~P. and {Donath}, A. and {Droettboom}, M. and {Earl}, N. and {Erben}, T. and {Fabbro}, S. and {Ferreira}, L.~A. and {Finethy}, T. and {Fox}, R.~T. and {Garrison}, L.~H. and {Gibbons}, S.~L.~J. and {Goldstein}, D.~A. and {Gommers}, R. and {Greco}, J.~P. and {Greenfield}, P. and {Groener}, A.~M. and {Grollier}, F. and {Hagen}, A. and {Hirst}, P. and {Homeier}, D. and {Horton}, A.~J. and {Hosseinzadeh}, G. and {Hu}, L. and {Hunkeler}, J.~S. and {Ivezi{\'c}}, {\v{Z}}. and {Jain}, A. and {Jenness}, T. and {Kanarek}, G. and {Kendrew}, S. and {Kern}, N.~S. and {Kerzendorf}, W.~E. and {Khvalko}, A. and {King}, J. and {Kirkby}, D. and {Kulkarni}, A.~M. and {Kumar}, A. and {Lee}, A. and {Lenz}, D. and {Littlefair}, S.~P. and {Ma}, Z. and {Macleod}, D.~M. and {Mastropietro}, M. and {McCully}, C. and {Montagnac}, S. and {Morris}, B.~M. and {Mueller}, M. and {Mumford}, S.~J. and {Muna}, D. and {Murphy}, N.~A. and {Nelson}, S. and {Nguyen}, G.~H. and {Ninan}, J.~P. and {N{\"o}the}, M. and {Ogaz}, S. and {Oh}, S. and {Parejko}, J.~K. and {Parley}, N. and {Pascual}, S. and {Patil}, R. and {Patil}, A.~A. and {Plunkett}, A.~L. and {Prochaska}, J.~X. and {Rastogi}, T. and {Reddy Janga}, V. and {Sabater}, J. and {Sakurikar}, P. and {Seifert}, M. and {Sherbert}, L.~E. and {Sherwood-Taylor}, H. and {Shih}, A.~Y. and {Sick}, J. and {Silbiger}, M.~T. and {Singanamalla}, S. and {Singer}, L.~P. and {Sladen}, P.~H. and {Sooley}, K.~A. and {Sornarajah}, S. and {Streicher}, O. and {Teuben}, P. and {Thomas}, S.~W. and {Tremblay}, G.~R. and {Turner}, J.~E.~H. and {Terr{\'o}n}, V. and {van Kerkwijk}, M.~H. and {de la Vega}, A. and {Watkins}, L.~L. and {Weaver}, B.~A. and {Whitmore}, J.~B. and {Woillez}, J. and {Zabalza}, V. and {Astropy Contributors}},
        title = "{The Astropy Project: Building an Open-science Project and Status of the v2.0 Core Package}",
      journal = {\aj},
         year = 2018,
        month = sep,
       volume = {156},
       number = {3},
          eid = {123},
        pages = {123},
          doi = {10.3847/1538-3881/aabc4f},
archivePrefix = {arXiv},
       eprint = {1801.02634},
 primaryClass = {astro-ph.IM},
       adsurl = {https://ui.adsabs.harvard.edu/abs/2018AJ....156..123A}
}

@ARTICLE{2013A&A...558A..33A,
       author = {{Astropy Collaboration} and {Robitaille}, Thomas P. and
         {Tollerud}, Erik J. and {Greenfield}, Perry and {Droettboom}, Michael and
         {Bray}, Erik and {Aldcroft}, Tom and {Davis}, Matt and
         {Ginsburg}, Adam and {Price-Whelan}, Adrian M. and
         {Kerzendorf}, Wolfgang E. and {Conley}, Alexander and {Crighton}, Neil and
         {Barbary}, Kyle and {Muna}, Demitri and {Ferguson}, Henry and
         {Grollier}, Fr{\'e}d{\'e}ric and {Parikh}, Madhura M. and
         {Nair}, Prasanth H. and {Unther}, Hans M. and {Deil}, Christoph and
         {Woillez}, Julien and {Conseil}, Simon and {Kramer}, Roban and
         {Turner}, James E.~H. and {Singer}, Leo and {Fox}, Ryan and
         {Weaver}, Benjamin A. and {Zabalza}, Victor and {Edwards}, Zachary I. and
         {Azalee Bostroem}, K. and {Burke}, D.~J. and {Casey}, Andrew R. and
         {Crawford}, Steven M. and {Dencheva}, Nadia and {Ely}, Justin and
         {Jenness}, Tim and {Labrie}, Kathleen and {Lim}, Pey Lian and
         {Pierfederici}, Francesco and {Pontzen}, Andrew and {Ptak}, Andy and
         {Refsdal}, Brian and {Servillat}, Mathieu and {Streicher}, Ole},
        title = "{Astropy: A community Python package for astronomy}",
      journal = {\aap},
         year = "2013",
        month = "Oct",
       volume = {558},
          eid = {A33},
        pages = {A33},
          doi = {10.1051/0004-6361/201322068},
archivePrefix = {arXiv},
       eprint = {1307.6212},
 primaryClass = {astro-ph.IM},
       adsurl = {https://ui.adsabs.harvard.edu/abs/2013A&A...558A..33A}
}

@ARTICLE{2010ApJS..188..405A,
       author = {{Abdo}, A.~A. and {Ackermann}, M. and {Ajello}, M. and {Allafort}, A. and {Antolini}, E. and {Atwood}, W.~B. and {Axelsson}, M. and {Baldini}, L. and {Ballet}, J. and {Barbiellini}, G. and et al.},
        title = "{Fermi Large Area Telescope First Source Catalog}",
      journal = {\apjs},
         year = 2010,
        month = jun,
       volume = {188},
       number = {2},
        pages = {405-436},
          doi = {10.1088/0067-0049/188/2/405},
archivePrefix = {arXiv},
       eprint = {1002.2280},
 primaryClass = {astro-ph.HE},
       adsurl = {https://ui.adsabs.harvard.edu/abs/2010ApJS..188..405A}
}

@ARTICLE{2022ApJS..260...53A,
       author = {{Abdollahi}, S. and {Acero}, F. and {Baldini}, L. and {Ballet}, J. and {Bastieri}, D. and {Bellazzini}, R. and {Berenji}, B. and {Berretta}, A. and {Bissaldi}, E. and {Blandford}, R.~D. and {Bloom}, E. and {Bonino}, R. and {Brill}, A. and {Britto}, R.~J. and {Bruel}, P. and {Burnett}, T.~H. and {Buson}, S. and {Cameron}, R.~A. and {Caputo}, R. and {Caraveo}, P.~A. and {Castro}, D. and {Chaty}, S. and {Cheung}, C.~C. and {Chiaro}, G. and {Cibrario}, N. and {Ciprini}, S. and {Coronado-Bl{\'a}zquez}, J. and {Crnogorcevic}, M. and {Cutini}, S. and {D'Ammando}, F. and {De Gaetano}, S. and {Digel}, S.~W. and {Di Lalla}, N. and {Dirirsa}, F. and {Di Venere}, L. and {Dom{\'\i}nguez}, A. and {Fallah Ramazani}, V. and {Fegan}, S.~J. and {Ferrara}, E.~C. and {Fiori}, A. and {Fleischhack}, H. and {Franckowiak}, A. and {Fukazawa}, Y. and {Funk}, S. and {Fusco}, P. and {Galanti}, G. and {Gammaldi}, V. and {Gargano}, F. and {Garrappa}, S. and {Gasparrini}, D. and {Giacchino}, F. and {Giglietto}, N. and {Giordano}, F. and {Giroletti}, M. and {Glanzman}, T. and {Green}, D. and {Grenier}, I.~A. and {Grondin}, M. -H. and {Guillemot}, L. and {Guiriec}, S. and {Gustafsson}, M. and {Harding}, A.~K. and {Hays}, E. and {Hewitt}, J.~W. and {Horan}, D. and {Hou}, X. and {J{\'o}hannesson}, G. and {Karwin}, C. and {Kayanoki}, T. and {Kerr}, M. and {Kuss}, M. and {Landriu}, D. and {Larsson}, S. and {Latronico}, L. and {Lemoine-Goumard}, M. and {Li}, J. and {Liodakis}, I. and {Longo}, F. and {Loparco}, F. and {Lott}, B. and {Lubrano}, P. and {Maldera}, S. and {Malyshev}, D. and {Manfreda}, A. and {Mart{\'\i}-Devesa}, G. and {Mazziotta}, M.~N. and {Mereu}, I. and {Meyer}, M. and {Michelson}, P.~F. and {Mirabal}, N. and {Mitthumsiri}, W. and {Mizuno}, T. and {Moiseev}, A.~A. and {Monzani}, M.~E. and {Morselli}, A. and {Moskalenko}, I.~V. and {Negro}, M. and {Nuss}, E. and {Omodei}, N. and {Orienti}, M. and {Orlando}, E. and {Paneque}, D. and {Pei}, Z. and {Perkins}, J.~S. and {Persic}, M. and {Pesce-Rollins}, M. and {Petrosian}, V. and {Pillera}, R. and {Poon}, H. and {Porter}, T.~A. and {Principe}, G. and {Rain{\`o}}, S. and {Rando}, R. and {Rani}, B. and {Razzano}, M. and {Razzaque}, S. and {Reimer}, A. and {Reimer}, O. and {Reposeur}, T. and {S{\'a}nchez-Conde}, M. and {Saz Parkinson}, P.~M. and {Scotton}, L. and {Serini}, D. and {Sgr{\`o}}, C. and {Siskind}, E.~J. and {Smith}, D.~A. and {Spandre}, G. and {Spinelli}, P. and {Sueoka}, K. and {Suson}, D.~J. and {Tajima}, H. and {Tak}, D. and {Thayer}, J.~B. and {Thompson}, D.~J. and {Torres}, D.~F. and {Troja}, E. and {Valverde}, J. and {Wood}, K. and {Zaharijas}, G.},
        title = "{Incremental Fermi Large Area Telescope Fourth Source Catalog}",
      journal = {\apjs},
         year = 2022,
        month = jun,
       volume = {260},
       number = {2},
          eid = {53},
        pages = {53},
          doi = {10.3847/1538-4365/ac6751},
archivePrefix = {arXiv},
       eprint = {2201.11184},
 primaryClass = {astro-ph.HE},
       adsurl = {https://ui.adsabs.harvard.edu/abs/2022ApJS..260...53A}
}

@ARTICLE{2016ApJ...824L..20A,
       author = {{Ackermann}, M. and {Anantua}, R. and {Asano}, K. and {Baldini}, L. and {Barbiellini}, G. and {Bastieri}, D. and {Becerra Gonzalez}, J. and {Bellazzini}, R. and {Bissaldi}, E. and {Blandford}, R.~D. and {Bloom}, E.~D. and {Bonino}, R. and {Bottacini}, E. and {Bruel}, P. and {Buehler}, R. and {Caliandro}, G.~A. and {Cameron}, R.~A. and {Caragiulo}, M. and {Caraveo}, P.~A. and {Cavazzuti}, E. and {Cecchi}, C. and {Cheung}, C.~C. and {Chiang}, J. and {Chiaro}, G. and {Ciprini}, S. and {Cohen-Tanugi}, J. and {Costanza}, F. and {Cutini}, S. and {D'Ammando}, F. and {de Palma}, F. and {Desiante}, R. and {Digel}, S.~W. and {Di Lalla}, N. and {Di Mauro}, M. and {Di Venere}, L. and {Drell}, P.~S. and {Favuzzi}, C. and {Fegan}, S.~J. and {Ferrara}, E.~C. and {Fukazawa}, Y. and {Funk}, S. and {Fusco}, P. and {Gargano}, F. and {Gasparrini}, D. and {Giglietto}, N. and {Giordano}, F. and {Giroletti}, M. and {Grenier}, I.~A. and {Guillemot}, L. and {Guiriec}, S. and {Hayashida}, M. and {Hays}, E. and {Horan}, D. and {J{\'o}hannesson}, G. and {Kensei}, S. and {Kocevski}, D. and {Kuss}, M. and {La Mura}, G. and {Larsson}, S. and {Latronico}, L. and {Li}, J. and {Longo}, F. and {Loparco}, F. and {Lott}, B. and {Lovellette}, M.~N. and {Lubrano}, P. and {Madejski}, G.~M. and {Magill}, J.~D. and {Maldera}, S. and {Manfreda}, A. and {Mayer}, M. and {Mazziotta}, M.~N. and {Michelson}, P.~F. and {Mirabal}, N. and {Mizuno}, T. and {Monzani}, M.~E. and {Morselli}, A. and {Moskalenko}, I.~V. and {Nalewajko}, K. and {Negro}, M. and {Nuss}, E. and {Ohsugi}, T. and {Orlando}, E. and {Paneque}, D. and {Perkins}, J.~S. and {Pesce-Rollins}, M. and {Piron}, F. and {Pivato}, G. and {Porter}, T.~A. and {Principe}, G. and {Rando}, R. and {Razzano}, M. and {Razzaque}, S. and {Reimer}, A. and {Scargle}, J.~D. and {Sgr{\`o}}, C. and {Sikora}, M. and {Simone}, D. and {Siskind}, E.~J. and {Spada}, F. and {Spinelli}, P. and {Stawarz}, L. and {Thayer}, J.~B. and {Thompson}, D.~J. and {Torres}, D.~F. and {Troja}, E. and {Uchiyama}, Y. and {Yuan}, Y. and {Zimmer}, S.},
        title = "{Minute-timescale >100 MeV {\ensuremath{\gamma}}-Ray Variability during the Giant Outburst of Quasar 3C 279 Observed by Fermi-LAT in 2015 June}",
      journal = {\apjl},
         year = 2016,
        month = jun,
       volume = {824},
       number = {2},
          eid = {L20},
        pages = {L20},
          doi = {10.3847/2041-8205/824/2/L20},
archivePrefix = {arXiv},
       eprint = {1605.05324},
 primaryClass = {astro-ph.HE},
       adsurl = {https://ui.adsabs.harvard.edu/abs/2016ApJ...824L..20A}
}

@INPROCEEDINGS{1997ASSL..218..163A,
       author = {{Alexander}, Tal},
        title = "{Is AGN Variability Correlated with Other AGN Properties? ZDCF Analysis of Small Samples of Sparse Light Curves}",
    booktitle = {Astronomical Time Series},
         year = 1997,
       editor = {{Maoz}, D. and {Sternberg}, A. and {Leibowitz}, E.~M.},
       series = {Astrophysics and Space Science Library},
       volume = {218},
        month = jan,
        pages = {163},
          doi = {10.1007/978-94-015-8941-3_14},
       adsurl = {https://ui.adsabs.harvard.edu/abs/1997ASSL..218..163A}
}

@ARTICLE{2021ApJ...906....5A,
       author = {{Amaya-Almaz{\'a}n}, Ra{\'u}l A. and {Chavushyan}, Vahram and {Pati{\~n}o-{\'A}lvarez}, Victor M.},
        title = "{Multiwavelength Analysis and the Difference in the Behavior of the Spectral Features during the 2010 and 2014 Flaring Periods of the Blazar 3C 454.3}",
      journal = {\apj},
         year = 2021,
        month = jan,
       volume = {906},
       number = {1},
          eid = {5},
        pages = {5},
          doi = {10.3847/1538-4357/abc689},
archivePrefix = {arXiv},
       eprint = {2010.13224},
 primaryClass = {astro-ph.HE},
       adsurl = {https://ui.adsabs.harvard.edu/abs/2021ApJ...906....5A}
}

@ARTICLE{1980ARA&A..18..321A,
       author = {{Angel}, J.~R.~P. and {Stockman}, H.~S.},
        title = "{Optical and infrared polarization of active extragalactic objects}",
      journal = {\araa},
         year = 1980,
        month = jan,
       volume = {18},
        pages = {321-361},
          doi = {10.1146/annurev.aa.18.090180.001541},
       adsurl = {https://ui.adsabs.harvard.edu/abs/1980ARA&A..18..321A}
}

@ARTICLE{2021MNRAS.504.5074S,
       author = {{Sahakyan}, N.},
        title = "{Modelling the broad-band emission of 3C 454.3}",
      journal = {\mnras},
         year = 2021,
        month = jul,
       volume = {504},
       number = {4},
        pages = {5074-5086},
          doi = {10.1093/mnras/stab1135},
archivePrefix = {arXiv},
       eprint = {2104.09831},
 primaryClass = {astro-ph.HE},
       adsurl = {https://ui.adsabs.harvard.edu/abs/2021MNRAS.504.5074S}
}

@ARTICLE{1996ApJ...461..657B,
       author = {{Bloom}, Steven D. and {Marscher}, Alan P.},
        title = "{An Analysis of the Synchrotron Self-Compton Model for the Multi--Wave Band Spectra of Blazars}",
      journal = {\apj},
         year = 1996,
        month = apr,
       volume = {461},
        pages = {657},
          doi = {10.1086/177092},
       adsurl = {https://ui.adsabs.harvard.edu/abs/1996ApJ...461..657B}
}

@ARTICLE{2020ApJ...891...68C,
       author = {{Chavushyan}, Vahram and {Pati{\~n}o-{\'A}lvarez}, Victor M. and {Amaya-Almaz{\'a}n}, Ra{\'u}l A. and {Carrasco}, Luis},
        title = "{Flare-like Variability of the Mg II {\ensuremath{\lambda}}2798 {\r{A}} Emission Line and UV Fe II Band in the Blazar CTA 102}",
      journal = {\apj},
         year = 2020,
        month = mar,
       volume = {891},
       number = {1},
          eid = {68},
        pages = {68},
          doi = {10.3847/1538-4357/ab6ef6},
archivePrefix = {arXiv},
       eprint = {2001.08296},
 primaryClass = {astro-ph.HE},
       adsurl = {https://ui.adsabs.harvard.edu/abs/2020ApJ...891...68C}
}

@ARTICLE{2011ApJ...736L..38V,
       author = {{Vercellone}, S. and {Striani}, E. and {Vittorini}, V. and {Donnarumma}, I. and {Pacciani}, L. and {Pucella}, G. and {Tavani}, M. and {Raiteri}, C.~M. and {Villata}, M. and {Romano}, P. and {Fiocchi}, M. and {Bazzano}, A. and {Bianchin}, V. and {Ferrigno}, C. and {Maraschi}, L. and {Pian}, E. and {T{\"u}rler}, M. and {Ubertini}, P. and {Bulgarelli}, A. and {Chen}, A.~W. and {Giuliani}, A. and {Longo}, F. and {Barbiellini}, G. and {Cardillo}, M. and {Cattaneo}, P.~W. and {Del Monte}, E. and {Evangelista}, Y. and {Feroci}, M. and {Ferrari}, A. and {Fuschino}, F. and {Gianotti}, F. and {Giusti}, M. and {Lazzarotto}, F. and {Pellizzoni}, A. and {Piano}, G. and {Pilia}, M. and {Rapisarda}, M. and {Rappoldi}, A. and {Sabatini}, S. and {Soffitta}, P. and {Trifoglio}, M. and {Trois}, A. and {Giommi}, P. and {Lucarelli}, F. and {Pittori}, C. and {Santolamazza}, P. and {Verrecchia}, F. and {Agudo}, I. and {Aller}, H.~D. and {Aller}, M.~F. and {Arkharov}, A.~A. and {Bach}, U. and {Berdyugin}, A. and {Borman}, G.~A. and {Chigladze}, R. and {Efimov}, Yu. S. and {Efimova}, N.~V. and {G{\'o}mez}, J.~L. and {Gurwell}, M.~A. and {McHardy}, I.~M. and {Joshi}, M. and {Kimeridze}, G.~N. and {Krajci}, T. and {Kurtanidze}, O.~M. and {Kurtanidze}, S.~O. and {Larionov}, V.~M. and {Lindfors}, E. and {Molina}, S.~N. and {Morozova}, D.~A. and {Nazarov}, S.~V. and {Nikolashvili}, M.~G. and {Nilsson}, K. and {Pasanen}, M. and {Reinthal}, R. and {Ros}, J.~A. and {Sadun}, A.~C. and {Sakamoto}, T. and {Sallum}, S. and {Sergeev}, S.~G. and {Schwartz}, R.~D. and {Sigua}, L.~A. and {Sillanp{\"a}{\"a}}, A. and {Sokolovsky}, K.~V. and {Strelnitski}, V. and {Takalo}, L. and {Taylor}, B. and {Walker}, G.},
        title = "{The Brightest Gamma-Ray Flaring Blazar in the Sky: AGILE and Multi-wavelength Observations of 3C 454.3 During 2010 November}",
      journal = {\apjl},
         year = 2011,
        month = aug,
       volume = {736},
       number = {2},
          eid = {L38},
        pages = {L38},
          doi = {10.1088/2041-8205/736/2/L38},
archivePrefix = {arXiv},
       eprint = {1106.5162},
 primaryClass = {astro-ph.HE},
       adsurl = {https://ui.adsabs.harvard.edu/abs/2011ApJ...736L..38V}
}

@ARTICLE{2017A&ARv..25....4B,
       author = {{Boccardi}, B. and {Krichbaum}, T.~P. and {Ros}, E. and {Zensus}, J.~A.},
        title = "{Radio observations of active galactic nuclei with mm-VLBI}",
      journal = {\aapr},
         year = 2017,
        month = nov,
       volume = {25},
       number = {1},
          eid = {4},
        pages = {4},
          doi = {10.1007/s00159-017-0105-6},
archivePrefix = {arXiv},
       eprint = {1711.07548},
 primaryClass = {astro-ph.HE},
       adsurl = {https://ui.adsabs.harvard.edu/abs/2017A&ARv..25....4B}
}

@ARTICLE{2011ApJ...735L..10A,
       author = {{Agudo}, Iv{\'a}n and {Marscher}, Alan P. and {Jorstad}, Svetlana G. and {Larionov}, Valeri M. and {G{\'o}mez}, Jos{\'e} L. and {L{\"a}hteenm{\"a}ki}, Anne and {Smith}, Paul S. and {Nilsson}, Kari and {Readhead}, Anthony C.~S. and {Aller}, Margo F. and {Heidt}, Jochen and {Gurwell}, Mark and {Thum}, Clemens and {Wehrle}, Ann E. and {Nikolashvili}, Maria G. and {Aller}, Hugh D. and {Ben{\'\i}tez}, Erika and {Blinov}, Dmitriy A. and {Hagen-Thorn}, Vladimir A. and {Hiriart}, David and {Jannuzi}, Buell T. and {Joshi}, Manasvita and {Kimeridze}, Givi N. and {Kurtanidze}, Omar M. and {Kurtanidze}, Sofia O. and {Lindfors}, Elina and {Molina}, Sol N. and {Morozova}, Daria A. and {Nieppola}, Elina and {Olmstead}, Alice R. and {Reinthal}, Riho and {Roca-Sogorb}, Mar and {Schmidt}, Gary D. and {Sigua}, Lorand A. and {Sillanp{\"a}{\"a}}, Aimo and {Takalo}, Leo and {Taylor}, Brian and {Tornikoski}, Merja and {Troitsky}, Ivan S. and {Zook}, Alma C. and {Wiesemeyer}, Helmut},
        title = "{On the Location of the {\ensuremath{\gamma}}-Ray Outburst Emission in the BL Lacertae Object AO 0235+164 Through Observations Across the Electromagnetic Spectrum}",
      journal = {\apjl},
         year = 2011,
        month = jul,
       volume = {735},
       number = {1},
          eid = {L10},
        pages = {L10},
          doi = {10.1088/2041-8205/735/1/L10},
archivePrefix = {arXiv},
       eprint = {1105.0549},
 primaryClass = {astro-ph.HE},
       adsurl = {https://ui.adsabs.harvard.edu/abs/2011ApJ...735L..10A}
}

@ARTICLE{2019NewAR..8701541H,
       author = {{Hovatta}, Talvikki and {Lindfors}, Elina},
        title = "{Relativistic Jets of Blazars}",
      journal = {\nar},
         year = 2019,
        month = dec,
       volume = {87},
          eid = {101541},
        pages = {101541},
          doi = {10.1016/j.newar.2020.101541},
archivePrefix = {arXiv},
       eprint = {2003.06322},
 primaryClass = {astro-ph.HE},
       adsurl = {https://ui.adsabs.harvard.edu/abs/2019NewAR..8701541H}
}

@ARTICLE{2005AJ....130.1418J,
       author = {{Jorstad}, Svetlana G. and {Marscher}, Alan P. and {Lister}, Matthew L. and {Stirling}, Alastair M. and {Cawthorne}, Timothy V. and {Gear}, Walter K. and {G{\'o}mez}, Jos{\'e} L. and {Stevens}, Jason A. and {Smith}, Paul S. and {Forster}, James R. and {Robson}, E. Ian},
        title = "{Polarimetric Observations of 15 Active Galactic Nuclei at High Frequencies: Jet Kinematics from Bimonthly Monitoring with the Very Long Baseline Array}",
      journal = {\aj},
         year = 2005,
        month = oct,
       volume = {130},
       number = {4},
        pages = {1418-1465},
          doi = {10.1086/444593},
archivePrefix = {arXiv},
       eprint = {astro-ph/0502501},
 primaryClass = {astro-ph},
       adsurl = {https://ui.adsabs.harvard.edu/abs/2005AJ....130.1418J}
}

@ARTICLE{1993ApJ...416..458D,
       author = {{Dermer}, Charles D. and {Schlickeiser}, Reinhard},
        title = "{Model for the High-Energy Emission from Blazars}",
      journal = {\apj},
         year = 1993,
        month = oct,
       volume = {416},
        pages = {458},
          doi = {10.1086/173251},
       adsurl = {https://ui.adsabs.harvard.edu/abs/1993ApJ...416..458D}
}

@ARTICLE{1989MNRAS.236..341G,
       author = {{Ghisellini}, Gabriele},
        title = "{Synchrotron self Compton models for compact sources - The case of a steep power-law particle distribution}",
      journal = {\mnras},
         year = 1989,
        month = jan,
       volume = {236},
        pages = {341-351},
          doi = {10.1093/mnras/236.2.341},
       adsurl = {https://ui.adsabs.harvard.edu/abs/1989MNRAS.236..341G}
}

@ARTICLE{1994ApJ...421..153S,
       author = {{Sikora}, Marek and {Begelman}, Mitchell C. and {Rees}, Martin J.},
        title = "{Comptonization of Diffuse Ambient Radiation by a Relativistic Jet: The Source of Gamma Rays from Blazars?}",
      journal = {\apj},
         year = 1994,
        month = jan,
       volume = {421},
        pages = {153},
          doi = {10.1086/173633},
       adsurl = {https://ui.adsabs.harvard.edu/abs/1994ApJ...421..153S}
}

@ARTICLE{2018MNRAS.479.2037P,
       author = {{Pati{\~n}o-{\'A}lvarez}, V.~M. and {Fernandes}, S. and {Chavushyan}, V. and {L{\'o}pez-Rodr{\'\i}guez}, E. and {Le{\'o}n-Tavares}, J. and {Schlegel}, E.~M. and {Carrasco}, L. and {Vald{\'e}s}, J. and {Carrami{\~n}ana}, A.},
        title = "{Multiwavelength photometric and spectropolarimetric analysis of the FSRQ 3C 279}",
      journal = {\mnras},
         year = 2018,
        month = sep,
       volume = {479},
       number = {2},
        pages = {2037-2064},
          doi = {10.1093/mnras/sty1497},
archivePrefix = {arXiv},
       eprint = {1806.01693},
 primaryClass = {astro-ph.HE},
       adsurl = {https://ui.adsabs.harvard.edu/abs/2018MNRAS.479.2037P}
}

@ARTICLE{2001ApJ...556..738J,
       author = {{Jorstad}, Svetlana G. and {Marscher}, Alan P. and {Mattox}, John R. and {Aller}, Margo F. and {Aller}, Hugh D. and {Wehrle}, Ann E. and {Bloom}, Steven D.},
        title = "{Multiepoch Very Long Baseline Array Observations of EGRET-detected Quasars and BL Lacertae Objects: Connection between Superluminal Ejections and Gamma-Ray Flares in Blazars}",
      journal = {\apj},
         year = 2001,
        month = aug,
       volume = {556},
       number = {2},
        pages = {738-748},
          doi = {10.1086/321605},
archivePrefix = {arXiv},
       eprint = {astro-ph/0102012},
 primaryClass = {astro-ph},
       adsurl = {https://ui.adsabs.harvard.edu/abs/2001ApJ...556..738J}
}

@ARTICLE{2015MNRAS.452.1280R,
       author = {{Ramakrishnan}, V. and {Hovatta}, T. and {Nieppola}, E. and {Tornikoski}, M. and {L{\"a}hteenm{\"a}ki}, A. and {Valtaoja}, E.},
        title = "{Locating the {\ensuremath{\gamma}}-ray emission site in Fermi/LAT blazars from correlation analysis between 37 GHz radio and {\ensuremath{\gamma}}-ray light curves}",
      journal = {\mnras},
         year = 2015,
        month = sep,
       volume = {452},
       number = {2},
        pages = {1280-1294},
          doi = {10.1093/mnras/stv321},
archivePrefix = {arXiv},
       eprint = {1507.04287},
 primaryClass = {astro-ph.HE},
       adsurl = {https://ui.adsabs.harvard.edu/abs/2015MNRAS.452.1280R}
}

@ARTICLE{2003ApJ...590...95L,
       author = {{L{\"a}hteenm{\"a}ki}, A. and {Valtaoja}, E.},
        title = "{Testing of Inverse Compton Models for Active Galactic Nuclei with Gamma-Ray and Radio Observations}",
      journal = {\apj},
         year = 2003,
        month = jun,
       volume = {590},
       number = {1},
        pages = {95-108},
          doi = {10.1086/374883},
       adsurl = {https://ui.adsabs.harvard.edu/abs/2003ApJ...590...95L}
}

@ARTICLE{2014MNRAS.441.1899F,
       author = {{Fuhrmann}, L. and {Larsson}, S. and {Chiang}, J. and {Angelakis}, E. and {Zensus}, J.~A. and {Nestoras}, I. and {Krichbaum}, T. {\^A}. P. and {Ungerechts}, H. and {Sievers}, A. and {Pavlidou}, V. and {Readhead}, A.~C.~S. and {Max-Moerbeck}, W. and {Pearson}, T.~J.},
        title = "{Detection of significant cm to sub-mm band radio and {\ensuremath{\gamma}}-ray correlated variability in Fermi bright blazars}",
      journal = {\mnras},
         year = 2014,
        month = jul,
       volume = {441},
       number = {3},
        pages = {1899-1909},
          doi = {10.1093/mnras/stu540},
archivePrefix = {arXiv},
       eprint = {1403.4170},
 primaryClass = {astro-ph.HE},
       adsurl = {https://ui.adsabs.harvard.edu/abs/2014MNRAS.441.1899F}
}

@ARTICLE{2018MNRAS.475..368H,
       author = {{Hodgson}, Jeffrey A. and {Rani}, Bindu and {Lee}, Sang-Sung and {Algaba}, Juan Carlos and {Kino}, Motoki and {Trippe}, Sascha and {Park}, Jong-Ho and {Zhao}, Guang-Yao and {Byun}, Do-Young and {Kang}, Sincheol and {Kim}, Jae-Young and {Kim}, Jeong-Sook and {Kim}, Soon-Wook and {Miyazaki}, Atsushi and {Wajima}, Kiyoaki and {Oh}, Junghwan and {Kim}, Dae-won and {Gurwell}, Mark},
        title = "{KVN observations reveal multiple {\ensuremath{\gamma}}-ray emission regions in 3C 84?}",
      journal = {\mnras},
         year = 2018,
        month = mar,
       volume = {475},
       number = {1},
        pages = {368-378},
          doi = {10.1093/mnras/stx3041},
archivePrefix = {arXiv},
       eprint = {1802.02763},
 primaryClass = {astro-ph.HE},
       adsurl = {https://ui.adsabs.harvard.edu/abs/2018MNRAS.475..368H}
}

@ARTICLE{2011ApJS..195...19O,
       author = {{Ogle}, Patrick M. and {Wehrle}, Ann E. and {Balonek}, Thomas and {Gurwell}, Mark A.},
        title = "{Blazar 3C 454.3 in Outburst and Quiescence during 2005-2007: Two Variable Synchrotron Emission Peaks}",
      journal = {\apjs},
         year = 2011,
        month = aug,
       volume = {195},
       number = {2},
          eid = {19},
        pages = {19},
          doi = {10.1088/0067-0049/195/2/19},
archivePrefix = {arXiv},
       eprint = {1003.3642},
 primaryClass = {astro-ph.CO},
       adsurl = {https://ui.adsabs.harvard.edu/abs/2011ApJS..195...19O}
}

@ARTICLE{2016Galax...5....3V,
       author = {{Villica{\~n}a-Pedraza}, Ilhuiyolitzin and {Carreto-Parra}, Francisco and {Carrami{\~n}ana}, Alberto and {Saucedo-Morales}, Julio},
        title = "{Multifrequency Study of the Blazar 3C 454.3}",
      journal = {Galaxies},
         year = 2016,
        month = dec,
       volume = {5},
       number = {1},
          eid = {3},
        pages = {3},
          doi = {10.3390/galaxies5010003},
       adsurl = {https://ui.adsabs.harvard.edu/abs/2016Galax...5....3V}
}

@ARTICLE{2021ApJS..253...10F,
       author = {{Fan}, J.~H. and {Kurtanidze}, S.~O. and {Liu}, Y. and {Kurtanidze}, O.~M. and {Nikolashvili}, M.~G. and {Liu}, X. and {Zhang}, L.~X. and {Cai}, J.~T. and {Zhu}, J.~T. and {He}, S.~L. and {Yang}, W.~X. and {Yang}, J.~H. and {Gu}, M.~F. and {Luo}, G.~Y. and {Yuan}, Y.~H.},
        title = "{Optical Photometry of the Quasar 3C 454.3 during the Period 2006-2018 and the Long-term Periodicity Analysis}",
      journal = {\apjs},
         year = 2021,
        month = mar,
       volume = {253},
       number = {1},
          eid = {10},
        pages = {10},
          doi = {10.3847/1538-4365/abd32d},
       adsurl = {https://ui.adsabs.harvard.edu/abs/2021ApJS..253...10F}
}

@ARTICLE{2011A&A...531A..95F,
       author = {{Fromm}, C.~M. and {Perucho}, M. and {Ros}, E. and {Savolainen}, T. and {Lobanov}, A.~P. and {Zensus}, J.~A. and {Aller}, M.~F. and {Aller}, H.~D. and {Gurwell}, M.~A. and {L{\"a}hteenm{\"a}ki}, A.},
        title = "{Catching the radio flare in CTA 102. I. Light curve analysis}",
      journal = {\aap},
         year = 2011,
        month = jul,
       volume = {531},
          eid = {A95},
        pages = {A95},
          doi = {10.1051/0004-6361/201116857},
archivePrefix = {arXiv},
       eprint = {1105.5024},
 primaryClass = {astro-ph.CO},
       adsurl = {https://ui.adsabs.harvard.edu/abs/2011A&A...531A..95F}
}

@ARTICLE{2023A&A...669A..32P,
       author = {{Paraschos}, G.~F. and {Mpisketzis}, V. and {Kim}, J. -Y. and {Witzel}, G. and {Krichbaum}, T.~P. and {Zensus}, J.~A. and {Gurwell}, M.~A. and {L{\"a}hteenm{\"a}ki}, A. and {Tornikoski}, M. and {Kiehlmann}, S. and {Readhead}, A.~C.~S.},
        title = "{A multi-band study and exploration of the radio wave-{\ensuremath{\gamma}}-ray connection in 3C 84}",
      journal = {\aap},
         year = 2023,
        month = jan,
       volume = {669},
          eid = {A32},
        pages = {A32},
          doi = {10.1051/0004-6361/202244814},
archivePrefix = {arXiv},
       eprint = {2210.09795},
 primaryClass = {astro-ph.HE},
       adsurl = {https://ui.adsabs.harvard.edu/abs/2023A&A...669A..32P}
}

@ARTICLE{2014A&A...571A..16P,
       author = {{Planck Collaboration} and {Ade}, P.~A.~R. and {Aghanim}, N. and {Armitage-Caplan}, C. and {Arnaud}, M. and {Ashdown}, M. and {Atrio-Barandela}, F. and {Aumont}, J. and {Baccigalupi}, C. and {Banday}, A.~J. and et al.},
        title = "{Planck 2013 results. XVI. Cosmological parameters}",
      journal = {\aap},
         year = 2014,
        month = nov,
       volume = {571},
          eid = {A16},
        pages = {A16},
          doi = {10.1051/0004-6361/201321591},
archivePrefix = {arXiv},
       eprint = {1303.5076},
 primaryClass = {astro-ph.CO},
       adsurl = {https://ui.adsabs.harvard.edu/abs/2014A&A...571A..16P}
}

@INPROCEEDINGS{2007ASPC..375..234G,
       author = {{Gurwell}, M.~A. and {Peck}, A.~B. and {Hostler}, S.~R. and {Darrah}, M.~R. and {Katz}, C.~A.},
        title = "{Monitoring Phase Calibrators at Submillimeter Wavelengths}",
    booktitle = {From Z-Machines to ALMA: (Sub)Millimeter Spectroscopy of Galaxies},
         year = 2007,
       editor = {{Baker}, A.~J. and {Glenn}, J. and {Harris}, A.~I. and {Mangum}, J.~G. and {Yun}, M.~S.},
       series = {Astronomical Society of the Pacific Conference Series},
       volume = {375},
        month = oct,
        pages = {234},
       adsurl = {https://ui.adsabs.harvard.edu/abs/2007ASPC..375..234G}
}

@ARTICLE{2013ApJ...779..100I,
       author = {{Isler}, Jedidah C. and {Urry}, C.~M. and {Coppi}, P. and {Bailyn}, C. and {Chatterjee}, R. and {Fossati}, G. and {Bonning}, E.~W. and {Maraschi}, L. and {Buxton}, M.},
        title = "{A Time-resolved Study of the Broad-line Region in Blazar 3C 454.3}",
      journal = {\apj},
         year = 2013,
        month = dec,
       volume = {779},
       number = {2},
          eid = {100},
        pages = {100},
          doi = {10.1088/0004-637X/779/2/100},
archivePrefix = {arXiv},
       eprint = {1310.0817},
 primaryClass = {astro-ph.CO},
       adsurl = {https://ui.adsabs.harvard.edu/abs/2013ApJ...779..100I}
}

@ARTICLE{2010ApJ...715..362J,
       author = {{Jorstad}, Svetlana G. and {Marscher}, Alan P. and {Larionov}, Valeri M. and {Agudo}, Iv{\'a}n and {Smith}, Paul S. and {Gurwell}, Mark and {L{\"a}hteenm{\"a}ki}, Anne and {Tornikoski}, Merja and {Markowitz}, Alex and {Arkharov}, Arkadi A. and {Blinov}, Dmitry A. and {Chatterjee}, Ritaban and {D'Arcangelo}, Francesca D. and {Falcone}, Abe D. and {G{\'o}mez}, Jos{\'e} L. and {Hagen-Thorn}, Vladimir A. and {Jordan}, Brendan and {Kimeridze}, Givi N. and {Konstantinova}, Tatiana S. and {Kopatskaya}, Evgenia N. and {Kurtanidze}, Omar and {Larionova}, Elena G. and {Larionova}, Liudmilla V. and {McHardy}, Ian M. and {Melnichuk}, Daria A. and {Roca-Sogorb}, Mar and {Schmidt}, Gary D. and {Skiff}, Brian and {Taylor}, Brian and {Thum}, Clemens and {Troitsky}, Ivan S. and {Wiesemeyer}, Helmut},
        title = "{Flaring Behavior of the Quasar 3C 454.3 Across the Electromagnetic Spectrum}",
      journal = {\apj},
         year = 2010,
        month = may,
       volume = {715},
       number = {1},
        pages = {362-384},
          doi = {10.1088/0004-637X/715/1/362},
archivePrefix = {arXiv},
       eprint = {1003.4293},
 primaryClass = {astro-ph.HE},
       adsurl = {https://ui.adsabs.harvard.edu/abs/2010ApJ...715..362J}
}

@ARTICLE{2013ApJ...773..147J,
       author = {{Jorstad}, Svetlana G. and {Marscher}, Alan P. and {Smith}, Paul S. and {Larionov}, Valeri M. and {Agudo}, Iv{\'a}n and {Gurwell}, Mark and {Wehrle}, Ann E. and {L{\"a}hteenm{\"a}ki}, Anne and {Nikolashvili}, Maria G. and {Schmidt}, Gary D. and {Arkharov}, Arkady A. and {Blinov}, Dmitry A. and {Blumenthal}, Kelly and {Casadio}, Carolina and {Chigladze}, Revaz A. and {Efimova}, Natalia V. and {Eggen}, Joseph R. and {G{\'o}mez}, Jos{\'e} L. and {Grupe}, Dirk and {Hagen-Thorn}, Vladimir A. and {Joshi}, Manasvita and {Kimeridze}, Givi N. and {Konstantinova}, Tatiana S. and {Kopatskaya}, Evgenia N. and {Kurtanidze}, Omar M. and {Kurtanidze}, Sofia O. and {Larionova}, Elena G. and {Larionova}, Liudmilla V. and {Sigua}, Lorand A. and {MacDonald}, Nicholas R. and {Maune}, Jeremy D. and {McHardy}, Ian M. and {Miller}, H. Richard and {Molina}, Sol N. and {Morozova}, Daria A. and {Scott}, Terri and {Taylor}, Brian W. and {Tornikoski}, Merja and {Troitsky}, Ivan S. and {Thum}, Clemens and {Walker}, Gary and {Williamson}, Karen E. and {Sallum}, Stephanie and {Consiglio}, Santina and {Strelnitski}, Vladimir},
        title = "{A Tight Connection between Gamma-Ray Outbursts and Parsec-scale Jet Activity in the Quasar 3C 454.3}",
      journal = {\apj},
         year = 2013,
        month = aug,
       volume = {773},
       number = {2},
          eid = {147},
        pages = {147},
          doi = {10.1088/0004-637X/773/2/147},
archivePrefix = {arXiv},
       eprint = {1307.2522},
 primaryClass = {astro-ph.HE},
       adsurl = {https://ui.adsabs.harvard.edu/abs/2013ApJ...773..147J}
}

@ARTICLE{2017ApJ...846...98J,
       author = {{Jorstad}, Svetlana G. and {Marscher}, Alan P. and {Morozova}, Daria A. and {Troitsky}, Ivan S. and {Agudo}, Iv{\'a}n and {Casadio}, Carolina and {Foord}, Adi and {G{\'o}mez}, Jos{\'e} L. and {MacDonald}, Nicholas R. and {Molina}, Sol N. and {L{\"a}hteenm{\"a}ki}, Anne and {Tammi}, Joni and {Tornikoski}, Merja},
        title = "{Kinematics of Parsec-scale Jets of Gamma-Ray Blazars at 43 GHz within the VLBA-BU-BLAZAR Program}",
      journal = {\apj},
         year = 2017,
        month = sep,
       volume = {846},
       number = {2},
          eid = {98},
        pages = {98},
          doi = {10.3847/1538-4357/aa8407},
archivePrefix = {arXiv},
       eprint = {1711.03983},
 primaryClass = {astro-ph.GA},
       adsurl = {https://ui.adsabs.harvard.edu/abs/2017ApJ...846...98J}
}

@ARTICLE{2011A&A...532A.146L,
       author = {{Le{\'o}n-Tavares}, J. and {Valtaoja}, E. and {Tornikoski}, M. and {L{\"a}hteenm{\"a}ki}, A. and {Nieppola}, E.},
        title = "{The connection between gamma-ray emission and millimeter flares in Fermi/LAT blazars}",
      journal = {\aap},
         year = 2011,
        month = aug,
       volume = {532},
          eid = {A146},
        pages = {A146},
          doi = {10.1051/0004-6361/201116664},
archivePrefix = {arXiv},
       eprint = {1102.1290},
 primaryClass = {astro-ph.CO},
       adsurl = {https://ui.adsabs.harvard.edu/abs/2011A&A...532A.146L}
}

@ARTICLE{2010ApJ...721.1383A,
       author = {{Ackermann}, M. and {Ajello}, M. and {Baldini}, L. and {Ballet}, J. and {Barbiellini}, G. and {Bastieri}, D. and {Bechtol}, K. and {Bellazzini}, R. and {Berenji}, B. and {Blandford}, R.~D. and {Bonamente}, E. and {Borgland}, A.~W. and {Bregeon}, J. and {Brigida}, M. and {Bruel}, P. and {Buehler}, R. and {Burnett}, T.~H. and {Buson}, S. and {Caliandro}, G.~A. and {Cameron}, R.~A. and {Caraveo}, P.~A. and {Carrigan}, S. and {Casandjian}, J.~M. and {Cavazzuti}, E. and {Cecchi}, C. and {{\c{C}}elik}, {\"O}. and {Chekhtman}, A. and {Cheung}, C.~C. and {Chiang}, J. and {Ciprini}, S. and {Claus}, R. and {Cohen-Tanugi}, J. and {Corbel}, S. and {Cutini}, S. and {D'Ammando}, F. and {Dermer}, C.~D. and {de Angelis}, A. and {de Palma}, F. and {Digel}, S.~W. and {Silva}, E. do Couto e. and {Drell}, P.~S. and {Dubois}, R. and {Dumora}, D. and {Escande}, L. and {Favuzzi}, C. and {Fegan}, S.~J. and {Ferrara}, E.~C. and {Fuhrmann}, L. and {Fukazawa}, Y. and {Fusco}, P. and {Gargano}, F. and {Gasparrini}, D. and {Gehrels}, N. and {Germani}, S. and {Giebels}, B. and {Giglietto}, N. and {Giommi}, P. and {Giordano}, F. and {Giroletti}, M. and {Glanzman}, T. and {Godfrey}, G. and {Grenier}, I.~A. and {Grove}, J.~E. and {Guiriec}, S. and {Hadasch}, D. and {Hayashida}, M. and {Hays}, E. and {J{\'o}hannesson}, G. and {Johnson}, A.~S. and {Johnson}, W.~N. and {Kamae}, T. and {Katagiri}, H. and {Kataoka}, J. and {Kn{\"o}dlseder}, J. and {Kuss}, M. and {Lande}, J. and {Larsson}, S. and {Latronico}, L. and {Lee}, S. -H. and {Llena Garde}, M. and {Longo}, F. and {Loparco}, F. and {Lott}, B. and {Lubrano}, P. and {Madejski}, G.~M. and {Makeev}, A. and {Marchili}, N. and {Mazziotta}, M.~N. and {McEnery}, J.~E. and {Mehault}, J. and {Michelson}, P.~F. and {Mizuno}, T. and {Monte}, C. and {Monzani}, M.~E. and {Morselli}, A. and {Moskalenko}, I.~V. and {Murgia}, S. and {Nakamori}, T. and {Nalewajko}, K. and {Naumann-Godo}, M. and {Nolan}, P.~L. and {Norris}, J.~P. and {Nuss}, E. and {Ohsugi}, T. and {Okumura}, A. and {Omodei}, N. and {Orlando}, E. and {Ormes}, J.~F. and {Pelassa}, V. and {Pepe}, M. and {Pesce-Rollins}, M. and {Piron}, F. and {Porter}, T.~A. and {Rain{\`o}}, S. and {Rando}, R. and {Razzano}, M. and {Reimer}, A. and {Reimer}, O. and {Reyes}, L.~C. and {Ripken}, J. and {Ritz}, S. and {Roth}, M. and {Sadrozinski}, H.~F. -W. and {Sanchez}, D. and {Sander}, A. and {Scargle}, J.~D. and {Sgr{\`o}}, C. and {Sikora}, M. and {Siskind}, E.~J. and {Spandre}, G. and {Spinelli}, P. and {Strickman}, M.~S. and {Suson}, D.~J. and {Takahashi}, H. and {Takahashi}, T. and {Tanaka}, T. and {Tanaka}, Y. and {Thayer}, J.~B. and {Thayer}, J.~G. and {Thompson}, D.~J. and {Tibaldo}, L. and {Torres}, D.~F. and {Tosti}, G. and {Tramacere}, A. and {Usher}, T.~L. and {Vandenbroucke}, J. and {Vilchez}, N. and {Vitale}, V. and {Waite}, A.~P. and {Wang}, P. and {Wehrle}, A.~E. and {Winer}, B.~L. and {Yang}, Z. and {Ylinen}, T. and {Ziegler}, M.},
        title = "{Fermi Gamma-ray Space Telescope Observations of Gamma-ray Outbursts from 3C 454.3 in 2009 December and 2010 April}",
      journal = {\apj},
         year = 2010,
        month = oct,
       volume = {721},
       number = {2},
        pages = {1383-1396},
          doi = {10.1088/0004-637X/721/2/1383},
archivePrefix = {arXiv},
       eprint = {1007.0483},
 primaryClass = {astro-ph.HE},
       adsurl = {https://ui.adsabs.harvard.edu/abs/2010ApJ...721.1383A}
}

@ARTICLE{2011ApJ...733L..26A,
       author = {{Abdo}, A.~A. and {Ackermann}, M. and {Ajello}, M. and {Allafort}, A. and {Baldini}, L. and {Ballet}, J. and {Barbiellini}, G. and {Bastieri}, D. and {Bellazzini}, R. and {Berenji}, B. and {Blandford}, R.~D. and {Bloom}, E.~D. and {Bonamente}, E. and {Borgland}, A.~W. and {Bouvier}, A. and {Bregeon}, J. and {Brigida}, M. and {Bruel}, P. and {Buehler}, R. and {Buson}, S. and {Caliandro}, G.~A. and {Cameron}, R.~A. and {Caraveo}, P.~A. and {Casandjian}, J.~M. and {Cavazzuti}, E. and {Cecchi}, C. and {Charles}, E. and {Chekhtman}, A. and {Cheung}, C.~C. and {Chiang}, J. and {Ciprini}, S. and {Claus}, R. and {Conrad}, J. and {Cutini}, S. and {D'Ammando}, F. and {de Angelis}, A. and {de Palma}, F. and {Dermer}, C.~D. and {Digel}, S.~W. and {Silva}, E. do Couto e. and {Drell}, P.~S. and {Dubois}, R. and {Dumora}, D. and {Escande}, L. and {Favuzzi}, C. and {Fegan}, S.~J. and {Ferrara}, E.~C. and {Fortin}, P. and {Fukazawa}, Y. and {Fusco}, P. and {Gargano}, F. and {Gasparrini}, D. and {Gehrels}, N. and {Germani}, S. and {Giglietto}, N. and {Giommi}, P. and {Giordano}, F. and {Giroletti}, M. and {Glanzman}, T. and {Godfrey}, G. and {Grenier}, I.~A. and {Grove}, J.~E. and {Guiriec}, S. and {Hadasch}, D. and {Hayashida}, M. and {Hays}, E. and {Horan}, D. and {Itoh}, R. and {J{\'o}hannesson}, G. and {Johnson}, A.~S. and {Kamae}, T. and {Katagiri}, H. and {Kataoka}, J. and {Kn{\"o}dlseder}, J. and {Kuss}, M. and {Lande}, J. and {Larsson}, S. and {Latronico}, L. and {Lee}, S. -H. and {Longo}, F. and {Loparco}, F. and {Lott}, B. and {Lovellette}, M.~N. and {Lubrano}, P. and {Madejski}, G.~M. and {Makeev}, A. and {Mazziotta}, M.~N. and {McConville}, W. and {McEnery}, J.~E. and {Michelson}, P.~F. and {Mitthumsiri}, W. and {Mizuno}, T. and {Moiseev}, A.~A. and {Monte}, C. and {Monzani}, M.~E. and {Morselli}, A. and {Moskalenko}, I.~V. and {Murgia}, S. and {Naumann-Godo}, M. and {Nishino}, S. and {Nolan}, P.~L. and {Norris}, J.~P. and {Nuss}, E. and {Ohsugi}, T. and {Okumura}, A. and {Orlando}, E. and {Ormes}, J.~F. and {Paneque}, D. and {Pelassa}, V. and {Pesce-Rollins}, M. and {Pierbattista}, M. and {Piron}, F. and {Porter}, T.~A. and {Rain{\`o}}, S. and {Rando}, R. and {Razzaque}, S. and {Reimer}, A. and {Reimer}, O. and {Ritz}, S. and {Roth}, M. and {Sadrozinski}, H.~F. -W. and {Sanchez}, D. and {Scargle}, J.~D. and {Schalk}, T.~L. and {Sgr{\`o}}, C. and {Siskind}, E.~J. and {Smith}, P.~D. and {Spandre}, G. and {Spinelli}, P. and {Strickman}, M.~S. and {Takahashi}, H. and {Takahashi}, T. and {Tanaka}, T. and {Tanaka}, Y. and {Thayer}, J.~G. and {Thayer}, J.~B. and {Thompson}, D.~J. and {Tibaldo}, L. and {Torres}, D.~F. and {Tosti}, G. and {Tramacere}, A. and {Troja}, E. and {Vandenbroucke}, J. and {Vasileiou}, V. and {Vianello}, G. and {Vilchez}, N. and {Vitale}, V. and {Waite}, A.~P. and {Wang}, P. and {Winer}, B.~L. and {Wood}, K.~S. and {Yang}, Z. and {Ziegler}, M.},
        title = "{Fermi Gamma-ray Space Telescope Observations of the Gamma-ray Outburst from 3C454.3 in November 2010}",
      journal = {\apjl},
         year = 2011,
        month = jun,
       volume = {733},
       number = {2},
        pages = {L26},
          doi = {10.1088/2041-8205/733/2/L26},
archivePrefix = {arXiv},
       eprint = {1102.0277},
 primaryClass = {astro-ph.HE},
       adsurl = {https://ui.adsabs.harvard.edu/abs/2011ApJ...733L..26A}
}

@ARTICLE{2013ApJ...763L..36L,
       author = {{Le{\'o}n-Tavares}, J. and {Chavushyan}, V. and {Pati{\~n}o-{\'A}lvarez}, V. and {Valtaoja}, E. and {Arshakian}, T.~G. and {Popovi{\'c}}, L. {\v{C}}. and {Tornikoski}, M. and {Lobanov}, A. and {Carrami{\~n}ana}, A. and {Carrasco}, L. and {L{\"a}hteenm{\"a}ki}, A.},
        title = "{Flare-like Variability of the Mg II {\ensuremath{\lambda}}2800 Emission Line in the {\ensuremath{\Gamma}}-Ray Blazar 3C 454.3}",
      journal = {\apjl},
         year = 2013,
        month = feb,
       volume = {763},
       number = {2},
          eid = {L36},
        pages = {L36},
          doi = {10.1088/2041-8205/763/2/L36},
archivePrefix = {arXiv},
       eprint = {1301.3064},
 primaryClass = {astro-ph.HE},
       adsurl = {https://ui.adsabs.harvard.edu/abs/2013ApJ...763L..36L}
}

@ARTICLE{2020ApJ...902...61L,
       author = {{Liodakis}, I. and {Blinov}, D. and {Jorstad}, S.~G. and {Arkharov}, A.~A. and {Di Paola}, A. and {Efimova}, N.~V. and {Grishina}, T.~S. and {Kiehlmann}, S. and {Kopatskaya}, E.~N. and {Larionov}, V.~M. and {Larionova}, L.~V. and {Larionova}, E.~G. and {Marscher}, A.~P. and {Morozova}, D.~A. and {Nikiforova}, A.~A. and {Pavlidou}, V. and {Traianou}, E. and {Troitskaya}, Yu. V. and {Troitsky}, I.~S. and {Uemura}, M. and {Weaver}, Z.~R.},
        title = "{Two Flares with One Shock: The Interesting Case of 3C 454.3}",
      journal = {\apj},
         year = 2020,
        month = oct,
       volume = {902},
       number = {1},
          eid = {61},
        pages = {61},
          doi = {10.3847/1538-4357/abb1b8},
archivePrefix = {arXiv},
       eprint = {2008.08603},
 primaryClass = {astro-ph.HE},
       adsurl = {https://ui.adsabs.harvard.edu/abs/2020ApJ...902...61L}
}

@ARTICLE{2018ApJS..234...12L,
       author = {{Lister}, M.~L. and {Aller}, M.~F. and {Aller}, H.~D. and {Hodge}, M.~A. and {Homan}, D.~C. and {Kovalev}, Y.~Y. and {Pushkarev}, A.~B. and {Savolainen}, T.},
        title = "{MOJAVE. XV. VLBA 15 GHz Total Intensity and Polarization Maps of 437 Parsec-scale AGN Jets from 1996 to 2017}",
      journal = {\apjs},
         year = 2018,
        month = jan,
       volume = {234},
       number = {1},
          eid = {12},
        pages = {12},
          doi = {10.3847/1538-4365/aa9c44},
archivePrefix = {arXiv},
       eprint = {1711.07802},
 primaryClass = {astro-ph.GA},
       adsurl = {https://ui.adsabs.harvard.edu/abs/2018ApJS..234...12L}
}

@ARTICLE{2009AJ....137.3718L,
       author = {{Lister}, M.~L. and {Aller}, H.~D. and {Aller}, M.~F. and {Cohen}, M.~H. and {Homan}, D.~C. and {Kadler}, M. and {Kellermann}, K.~I. and {Kovalev}, Y.~Y. and {Ros}, E. and {Savolainen}, T. and {Zensus}, J.~A. and {Vermeulen}, R.~C.},
        title = "{MOJAVE: Monitoring of Jets in Active Galactic Nuclei with VLBA Experiments. V. Multi-Epoch VLBA Images}",
      journal = {\aj},
         year = 2009,
        month = mar,
       volume = {137},
       number = {3},
        pages = {3718-3729},
          doi = {10.1088/0004-6256/137/3/3718},
archivePrefix = {arXiv},
       eprint = {0812.3947},
 primaryClass = {astro-ph},
       adsurl = {https://ui.adsabs.harvard.edu/abs/2009AJ....137.3718L}
}

@ARTICLE{1985ApJ...298..114M,
       author = {{Marscher}, A.~P. and {Gear}, W.~K.},
        title = "{Models for high-frequency radio outbursts in extragalactic sources, with application to the early 1983 millimeter-to-infrared flare of 3C 273.}",
      journal = {\apj},
         year = 1985,
        month = nov,
       volume = {298},
        pages = {114-127},
          doi = {10.1086/163592},
       adsurl = {https://ui.adsabs.harvard.edu/abs/1985ApJ...298..114M}
}

@ARTICLE{2010ApJ...710L.126M,
       author = {{Marscher}, Alan P. and {Jorstad}, Svetlana G. and {Larionov}, Valeri M. and {Aller}, Margo F. and {Aller}, Hugh D. and {L{\"a}hteenm{\"a}ki}, Anne and {Agudo}, Iv{\'a}n and {Smith}, Paul S. and {Gurwell}, Mark and {Hagen-Thorn}, Vladimir A. and {Konstantinova}, Tatiana S. and {Larionova}, Elena G. and {Larionova}, Liudmila V. and {Melnichuk}, Daria A. and {Blinov}, Dmitry A. and {Kopatskaya}, Evgenia N. and {Troitsky}, Ivan S. and {Tornikoski}, Merja and {Hovatta}, Talvikki and {Schmidt}, Gary D. and {D'Arcangelo}, Francesca D. and {Bhattarai}, Dipesh and {Taylor}, Brian and {Olmstead}, Alice R. and {Manne-Nicholas}, Emily and {Roca-Sogorb}, Mar and {G{\'o}mez}, Jos{\'e} L. and {McHardy}, Ian M. and {Kurtanidze}, Omar and {Nikolashvili}, Maria G. and {Kimeridze}, Givi N. and {Sigua}, Lorand A.},
        title = "{Probing the Inner Jet of the Quasar PKS 1510-089 with Multi-Waveband Monitoring During Strong Gamma-Ray Activity}",
      journal = {\apjl},
         year = 2010,
        month = feb,
       volume = {710},
       number = {2},
        pages = {L126-L131},
          doi = {10.1088/2041-8205/710/2/L126},
archivePrefix = {arXiv},
       eprint = {1001.2574},
 primaryClass = {astro-ph.CO},
       adsurl = {https://ui.adsabs.harvard.edu/abs/2010ApJ...710L.126M}
}

@INPROCEEDINGS{2007ASPC..376..127M,
       author = {{McMullin}, J.~P. and {Waters}, B. and {Schiebel}, D. and {Young}, W. and {Golap}, K.},
        title = "{CASA Architecture and Applications}",
    booktitle = {Astronomical Data Analysis Software and Systems XVI},
         year = 2007,
       editor = {{Shaw}, R.~A. and {Hill}, F. and {Bell}, D.~J.},
       series = {Astronomical Society of the Pacific Conference Series},
       volume = {376},
        month = oct,
        pages = {127},
       adsurl = {https://ui.adsabs.harvard.edu/abs/2007ASPC..376..127M}
}

@ARTICLE{2019A&A...631A...4N,
       author = {{Nalewajko}, Krzysztof and {Gupta}, Alok C. and {Liao}, Mai and {Hryniewicz}, Krzysztof and {Gupta}, Maitrayee and {Gu}, Minfeng},
        title = "{Long-term optical spectroscopic variations in blazar 3C 454.3}",
      journal = {\aap},
         year = 2019,
        month = nov,
       volume = {631},
          eid = {A4},
        pages = {A4},
          doi = {10.1051/0004-6361/201935904},
archivePrefix = {arXiv},
       eprint = {1909.01357},
 primaryClass = {astro-ph.HE},
       adsurl = {https://ui.adsabs.harvard.edu/abs/2019A&A...631A...4N}
}

@ARTICLE{2019A&A...630A..56P,
       author = {{Pati{\~n}o-{\'A}lvarez}, V.~M. and {Dzib}, S.~A. and {Lobanov}, A. and {Chavushyan}, V.},
        title = "{Is there a non-stationary {\ensuremath{\gamma}}-ray emission zone 42 pc from the 3C 279 core?}",
      journal = {\aap},
         year = 2019,
        month = oct,
       volume = {630},
          eid = {A56},
        pages = {A56},
          doi = {10.1051/0004-6361/201834401},
archivePrefix = {arXiv},
       eprint = {1907.08314},
 primaryClass = {astro-ph.HE},
       adsurl = {https://ui.adsabs.harvard.edu/abs/2019A&A...630A..56P}
}

@ARTICLE{1987Natur.328..778P,
       author = {{Pauliny-Toth}, I.~I.~K. and {Porcas}, R.~W. and {Zensus}, J.~A. and {Kellermann}, K.~I. and {Wu}, S.~Y. and {Nicholson}, G.~D. and {Mantovani}, F.},
        title = "{Peculiar variations in the structure of the quasar 3C454.3}",
      journal = {\nat},
         year = 1987,
        month = aug,
       volume = {328},
       number = {6133},
        pages = {778-782},
          doi = {10.1038/328778a0},
       adsurl = {https://ui.adsabs.harvard.edu/abs/1987Natur.328..778P}
}

@ARTICLE{2018ApJ...856...99P,
       author = {{Pittori}, C. and {Lucarelli}, F. and {Verrecchia}, F. and {Raiteri}, C.~M. and {Villata}, M. and {Vittorini}, V. and {Tavani}, M. and {Puccetti}, S. and {Perri}, M. and {Donnarumma}, I. and {Vercellone}, S. and {Acosta-Pulido}, J.~A. and {Bachev}, R. and {Ben{\'\i}tez}, E. and {Borman}, G.~A. and {Carnerero}, M.~I. and {Carosati}, D. and {Chen}, W.~P. and {Ehgamberdiev}, Sh. A. and {Goded}, A. and {Grishina}, T.~S. and {Hiriart}, D. and {Hsiao}, H.~Y. and {Jorstad}, S.~G. and {Kimeridze}, G.~N. and {Kopatskaya}, E.~N. and {Kurtanidze}, O.~M. and {Kurtanidze}, S.~O. and {Larionov}, V.~M. and {Larionova}, L.~V. and {Marscher}, A.~P. and {Mirzaqulov}, D.~O. and {Morozova}, D.~A. and {Nilsson}, K. and {Samal}, M.~R. and {Sigua}, L.~A. and {Spassov}, B. and {Strigachev}, A. and {Takalo}, L.~O. and {Antonelli}, L.~A. and {Bulgarelli}, A. and {Cattaneo}, P. and {Colafrancesco}, S. and {Giommi}, P. and {Longo}, F. and {Morselli}, A. and {Paoletti}, F.},
        title = "{The Bright {\ensuremath{\gamma}}-ray Flare of 3C 279 in 2015 June: AGILE Detection and Multifrequency Follow-up Observations}",
      journal = {\apj},
         year = 2018,
        month = apr,
       volume = {856},
       number = {2},
          eid = {99},
        pages = {99},
          doi = {10.3847/1538-4357/aab1f9},
archivePrefix = {arXiv},
       eprint = {1803.07529},
 primaryClass = {astro-ph.HE},
       adsurl = {https://ui.adsabs.harvard.edu/abs/2018ApJ...856...99P}
}

@ARTICLE{2011ApJS..194...29R,
       author = {{Richards}, Joseph L. and {Max-Moerbeck}, Walter and {Pavlidou}, Vasiliki and {King}, Oliver G. and {Pearson}, Timothy J. and {Readhead}, Anthony C.~S. and {Reeves}, Rodrigo and {Shepherd}, Martin C. and {Stevenson}, Matthew A. and {Weintraub}, Lawrence C. and {Fuhrmann}, Lars and {Angelakis}, Emmanouil and {Zensus}, J. Anton and {Healey}, Stephen E. and {Romani}, Roger W. and {Shaw}, Michael S. and {Grainge}, Keith and {Birkinshaw}, Mark and {Lancaster}, Katy and {Worrall}, Diana M. and {Taylor}, Gregory B. and {Cotter}, Garret and {Bustos}, Ricardo},
        title = "{Blazars in the Fermi Era: The OVRO 40 m Telescope Monitoring Program}",
      journal = {\apjs},
         year = 2011,
        month = jun,
       volume = {194},
       number = {2},
          eid = {29},
        pages = {29},
          doi = {10.1088/0067-0049/194/2/29},
archivePrefix = {arXiv},
       eprint = {1011.3111},
 primaryClass = {astro-ph.CO},
       adsurl = {https://ui.adsabs.harvard.edu/abs/2011ApJS..194...29R}
}

@ARTICLE{2008ApJ...675...71S,
       author = {{Sikora}, Marek and {Moderski}, Rafa{\L} and {Madejski}, Greg M.},
        title = "{3C 454.3 Reveals the Structure and Physics of Its ``Blazar Zone''}",
      journal = {\apj},
         year = 2008,
        month = mar,
       volume = {675},
       number = {1},
        pages = {71-78},
          doi = {10.1086/526419},
archivePrefix = {arXiv},
       eprint = {0711.3524},
 primaryClass = {astro-ph},
       adsurl = {https://ui.adsabs.harvard.edu/abs/2008ApJ...675...71S}
}

@ARTICLE{1993A&A...269...67M,
       author = {{Mannheim}, K.},
        title = "{The proton blazar.}",
      journal = {\aap},
         year = 1993,
        month = mar,
       volume = {269},
        pages = {67-76},
          doi = {10.48550/arXiv.astro-ph/9302006},
archivePrefix = {arXiv},
       eprint = {astro-ph/9302006},
 primaryClass = {astro-ph},
       adsurl = {https://ui.adsabs.harvard.edu/abs/1993A&A...269...67M}
}

@ARTICLE{1992ApJ...397L...5M,
       author = {{Maraschi}, L. and {Ghisellini}, G. and {Celotti}, A.},
        title = "{A Jet Model for the Gamma-Ray--emitting Blazar 3C 279}",
      journal = {\apjl},
         year = 1992,
        month = sep,
       volume = {397},
        pages = {L5},
          doi = {10.1086/186531},
       adsurl = {https://ui.adsabs.harvard.edu/abs/1992ApJ...397L...5M}
}
\bibliographystyle{aa}

\end{document}